\documentclass[letterpaper,twocolumn,10pt]{article}
\usepackage{usenix2019_v3}

\usepackage{xspace, color}
\newcommand{\sysname}{\textsc{Miner}\xspace}

\newcommand{\sysnamepart}{\textsc{Miner\_part}\xspace}

\usepackage{booktabs}
\usepackage{tabularx}
\usepackage{array}
\usepackage{booktabs}
\usepackage{multirow}
\usepackage{caption}
\usepackage{subcaption}
\usepackage{stfloats}
\usepackage{listings}
\usepackage{setspace}
\usepackage{paralist}
\usepackage{footnote}
\usepackage{graphicx}
\usepackage{authblk}

\usepackage[shortlabels]{enumitem}
\newcommand{\KL}[1]{}
\newcommand{\LV}[1] {\textcolor{cyan}{#1}}

\newcommand{\ignore}[1]{}

\newcommand{\wqy}[1] {{\color{purple}#1}}
\newcolumntype{L}[1]{>{\raggedright\let\newline\\\arraybackslash\hspace{0pt}}m{#1}}
\newcolumntype{C}[1]{>{\centering\let\newline\\\arraybackslash\hspace{0pt}}m{#1}}
\newcolumntype{R}[1]{>{\raggedleft\let\newline\\\arraybackslash\hspace{0pt}}m{#1}}

\usepackage{lipsum}
\usepackage{marvosym}

\begin{document}
\title{\sysname: A Hybrid Data-Driven Approach for REST API Fuzzing} 

\author[$\dag$]{Chenyang Lyu}
\author[$\dag$]{Jiacheng Xu}
\author[$\dag,{\textrm{\Letter}}$]{Shouling Ji}
\author[$\dag$]{Xuhong Zhang}
\author[$\dag$]{Qinying Wang}
\author[$^\ast$]{Binbin Zhao}
\author[$\dag$]{Gaoning Pan}
\author[$\ddag$]{Wei Cao}
\author[$\ddag$]{Peng Chen}
\author[$^\ast$]{Raheem Beyah}

\affil[ ]{$^\dag${\small Zhejiang University},
$^\ast${\small Georgia Institute of Technology}, 
$^\ddag${\small Ant Group}
}

\maketitle
\newcommand\blfootnote[1]{%
\begingroup
\renewcommand\thefootnote{}\footnote{#1}%
\addtocounter{footnote}{-1}%
\endgroup
}
\blfootnote{Shouling Ji is the corresponding author.}

\begin{abstract}

In recent years, REST API fuzzing has emerged to explore errors on a cloud service. Its performance highly depends on the sequence construction and request generation. However, existing REST API fuzzers have trouble generating long sequences with well-constructed requests to trigger hard-to-reach states in a cloud service, 
which limits their performance of finding deep errors and security bugs. 
Further, they cannot find the specific errors 
caused by using undefined parameters during request generation. 
Therefore, in this paper, we propose a novel hybrid data-driven solution, named \emph{\sysname}, with three new designs working together to address the above limitations. First, \sysname collects the valid sequences whose requests pass the cloud service's checking as the templates, and assigns more executions to long sequence templates. 
Second, to improve the generation quality of requests in a sequence template, 
\sysname creatively leverages the state-of-the-art neural network model to predict key request parameters and provide them with appropriate parameter values. 
Third, \sysname implements a new data-driven security rule checker to capture the new kind of errors caused by undefined parameters. 
We evaluate \sysname against the state-of-the-art fuzzer RESTler on GitLab, Bugzilla, and WordPress via 11 REST APIs. The results demonstrate that the average pass rate of \sysname is 23.42\% higher than RESTler. \sysname finds 97.54\% more unique errors than RESTler on average and 142.86\% more reproducible errors after manual analysis. We have reported all the newly found errors, and 7 of them have been confirmed as logic bugs by the corresponding vendors. 

\end{abstract}

\section{Introduction}\label{sec:intro}



Cloud services have experienced significant growth 
over the last decade. 
They provide software-as-a-service applications that can be programmatically accessed through \textit{REST APIs}~\cite{fielding2000architectural}, without the requirement for local software installations. 
In a shared cloud service architecture, multiple users can access the same application through separate processes. 
In this manner, incorrect REST API access may result in a crash of a process, which can finally lead to cloud service collapse, broken access control, and private data leakage.

Therefore,  to explore potential errors of cloud services via REST APIs, 
early works~\cite{apifuzzer,tntfuzzer,qualys}
leverage previously-captured API traffic 
and manually-defined rules to generate testing requests. 
Recently, to better automatically infer the dependencies among request types and construct request sequences, REST API fuzzing is proposed to test cloud services~\cite{restler, atlidakis2020checking, godefroid2020differential, atlidakis2020pythia}. 
Typically, 
1) the generation strategies of a REST API fuzzer handle two main problems, i.e., 
how to construct a sequence template and how to generate each request in a template. 
After generation, the fuzzer sends the ready-to-use request sequences to test a target cloud service; 
2) For each request in a sequence, the cloud service first checks whether the request conforms to the syntax and semantics. 
If not, it returns a response in \texttt{40$\times$} Range. 
On the contrary, the cloud service performs a behavior according to the request, e.g., the cloud service deletes a resource when receiving a \texttt{DELETE} request. 
If the cloud service behaves normally, e.g., it successfully deletes the specified resource according to the request, it returns a response in \texttt{20$\times$} Range. 
Otherwise, the cloud service behaves abnormally, e.g., it tries to delete a non-existent resource and results in an error state. 
Then, it returns a response in \texttt{50$\times$} Range; 
3) The fuzzer analyzes the response to each request and infers the triggered states of a cloud service; 
4) The fuzzer collects the unique request sequences triggering error states, e.g., the sequences having responses in \texttt{50$\times$} Range or violating security rules. 
Following the above steps, a REST API  fuzzer explores reachable cloud service states with 
generated request sequences. 

However, we find that existing REST API fuzzers 
cannot efficiently generate long request sequences to test cloud services. 
As a result, existing fuzzers have trouble finding deep errors hidden in hard-to-reach states of cloud services. 
In particular, it is hard for existing fuzzers to find security bugs like use-after-free bugs and resource-hierarchy bugs as introduced in~\cite{atlidakis2020checking}, which need to be triggered with a request sequence of no less than 3 in length. 
The underlying reason is that security bugs on cloud services often require complex execution logic to trigger.  
We summarize two main reasons, corresponding to the sequence construction and request generation, that the state-of-the-art REST API fuzzers fail to generate long request sequences. 
1) Existing fuzzers often fail to extend request sequence templates. Their sequence extension process frequently abandons the constructed sequence templates and starts with an empty template, making it challenging to construct sequences of larger lengths; 
2) 
When generating a request in a sequence template, 
existing fuzzers randomly select values for the request's parameters. 
They cannot figure out which parameter should be mutated or what parameter value should be assigned, i.e., the \textit{key mutation} on a request. 
As a result, they have trouble 
generating {valid} requests 
that can pass a cloud service's checking.

To overcome the above two limitations and improve error discovery, 
in this paper, we present a hybrid data-driven fuzzing solution named \sysname with the following three important designs. 1) \emph{Length-orientated sequence construction:} \sysname leverages the historical data to guide the sequence generation. Specifically, \sysname collects the sequence templates, whose requests have been successfully generated and have passed the cloud service's checking in the past fuzzing process, as the initial templates of the sequence extension process. Then, \sysname leverages a probability function to assign more selection times to the sequence templates with large lengths in the extension process; 
2) \emph{Attention model-based request generation:} 
To improve request generation quality, for each valid request that passes the checking, \sysname first collects its parameters and the used values as the key mutations. 
Second, \sysname constructs 
an attention model, which is one of the state-of-the-art neural network models, 
to learn the implicit relationship among the key mutations. Then, \sysname uses the model to generate more diverse combinations of desirable key mutations, which can construct high-quality requests to pass the checking and explore different states of a cloud service; 
3) \emph{Request parameter violation checking:} \sysname implements an extra security rule checker to capture incorrect parameter usage errors. 
By using this checker, \sysname generates a new request with an undefined parameter to test a cloud service. 
For instance, \sysname can construct a \texttt{PUT} request containing the undefined parameter, which comes from a \texttt{POST} request.  
If the request gets a response in \texttt{50$\times$} Range, \sysname infers that an incorrect parameter usage error is triggered, and records the used request locally for further analysis. 
The above three core designs are compatible with each other, 
and work together in the fuzzing process to improve the fuzzing performance.

Our approach is a generic data-driven framework that does not require initial training data before fuzzing. Instead, it automatically collects training data during the fuzzing process. It can be applied to most REST API fuzzers, and improves their performance under the same experimental condition. 
In this paper, we implement a prototype of \sysname based on the state-of-the-art fuzzer \textit{RESTler}~\cite{restler}. 
Then, we evaluate the performance of \sysname and RESTler on 3 open-sourced cloud services GitLab~\cite{gitlab}, Bugzilla~\cite{bugzilla} and WordPress~\cite{wordpress} via 11 representative REST APIs. The average pass rate of \sysname is 23.42\%  higher than RESTler, which indicates that \sysname significantly improves the request generation 
to pass the cloud service's checking. 
The evaluation also shows that our designs significantly improve the sequence construction of \sysname, which assigns most executions to the sequences of length greater than 4. 
Furthermore, \sysname discovers 142.86\% more reproducible errors than RESTler, including 5 security bugs that access deleted resources. 

In summary, we make the following main contributions.

$\bullet$ 
We find that  REST API fuzzers 
have trouble generating long sequences containing high-quality requests and cannot find incorrect parameter usage errors. 
Motivated by our findings, 
we propose a new solution named \sysname with three novel data-driven designs, 
i.e., 1) leveraging the history data to guide the sequence construction, 2) improving the request generation with desirable parameter values by constructing an 
attention model for prediction, and 3) exploring a new kind of errors by
generating requests with undefined parameters. 

$\bullet$ 
We evaluate \sysname and RESTler on 3 open-sourced cloud services, 
and show the significant performance of \sysname on sequence construction, request generation and error discovery. 
Furthermore, we utilize the published bugs of GitLab as the ground truth 
to demonstrate that \sysname performs better than RESTler on serious bug discovery. 
The stepwise analysis shows the contribution of our designs to pass rate and error discovery. 
We conduct extensive analysis to show 
high code line coverage, low time overhead, and diverse  execution distribution of requests achieved by \sysname \footnote{We  open source \sysname at \emph{https://github.com/puppet-meteor/MINER} 
to facilitate future research on REST API fuzzing.}.

\vspace{-2mm}
\section{Background}\label{sec:background}




\subsection{REST API for Cloud Services}
Most cloud services can be accessed via REST APIs to provide various functionalities. 
In implementation, different types of requests can be sent to trigger different behaviors of a cloud service. 
For instance, a client 
1) can send a \texttt{GET}  request to GitLab, 
in order to get a list of visible projects with the selected parameters after authentication; 
2) can send a \texttt{POST}  request to create a new project with the specified parameters; 
3) is able to update the parameters of a specified project by sending a \texttt{PUT}  request; 
and 4) can delete a project with a \texttt{DELETE}  request. 
These behaviors can explore different states of a cloud service. 
Similarly, by automatically generating and sending request sequences via a cloud service's REST API, 
a fuzzer can explore errors hidden in different states.

\vspace{-2mm}
\subsection{REST API Fuzzing}\label{sec:restapi}

Two of the main challenges to fuzz  cloud services are as follows: 
How to construct sequences composed of different requests to trigger a series of behaviors and explore deep states of a cloud service, 
and how to construct high-quality requests to pass the cloud service's checking. 
Towards this, REST API fuzzing is proposed 
to explore errors hidden in the reachable execution states of a cloud service~\cite{restler, atlidakis2020checking, godefroid2020differential, atlidakis2020pythia}. 
A REST API fuzzer automatically generates request sequences to test a target cloud service, 
and is guided by the request responses.  
If a request in the generated sequence triggers a response in \texttt{50$\times$} Range, the fuzzer considers that an error is triggered and stores the sequence for future analysis. 
The main modules of a REST API fuzzer are as follows. 

\begin{figure*}[t]
    \setlength{\abovecaptionskip}{0.5cm}
    \setlength{\belowcaptionskip}{-0.5cm}
     \centering
     \begin{subfigure}[b]{0.23\textwidth}    
         \centering
         \includegraphics[width=1.53in,height=1.38in]{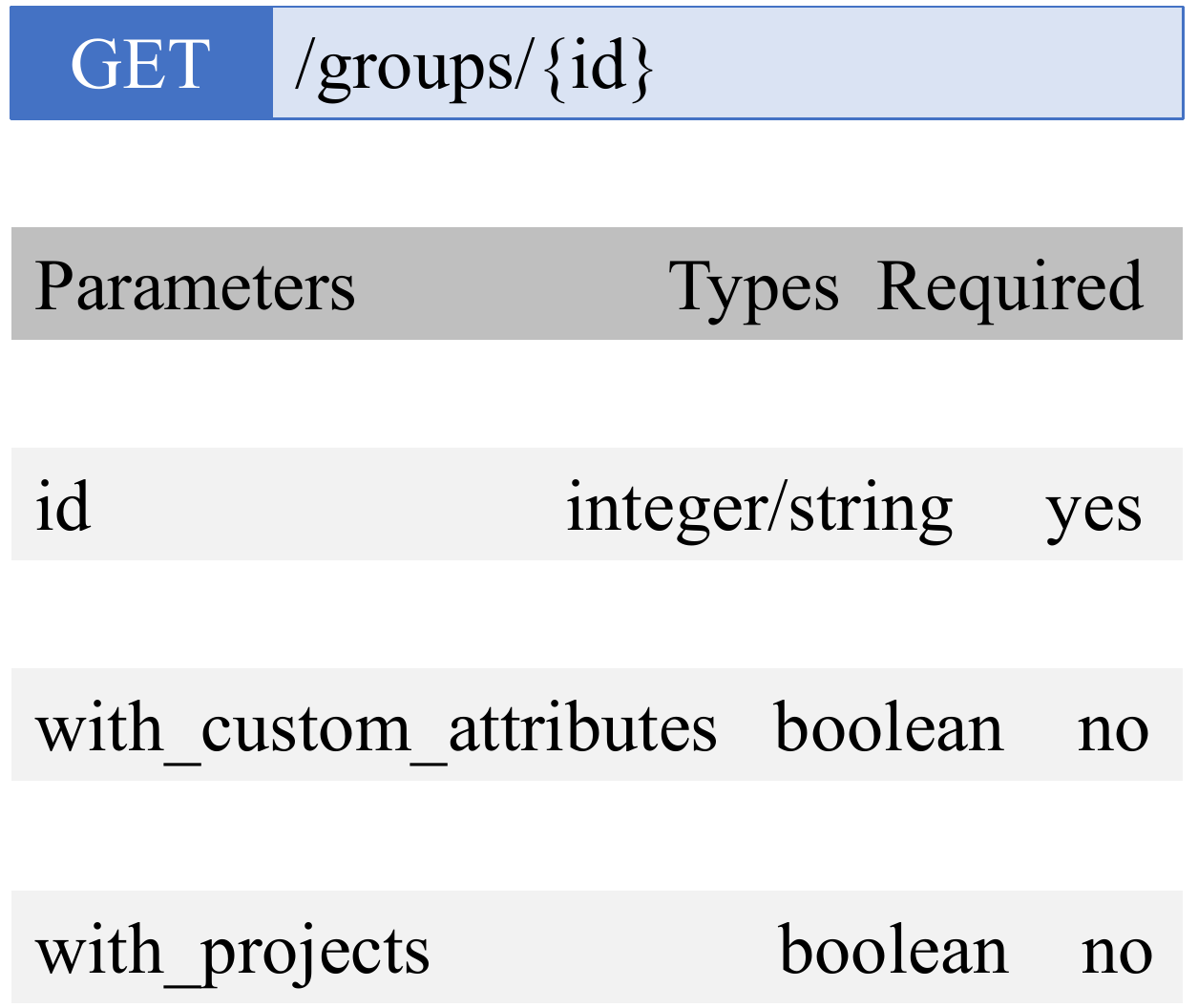}
         \caption{An example description of "GET /groups/\{id\}" on the web.}
         \label{fig:specflow1}
     \end{subfigure}
     \hfill
     \begin{subfigure}[b]{0.18\textwidth}    
         \centering
         \includegraphics[width=1.23in,height=1.38in]{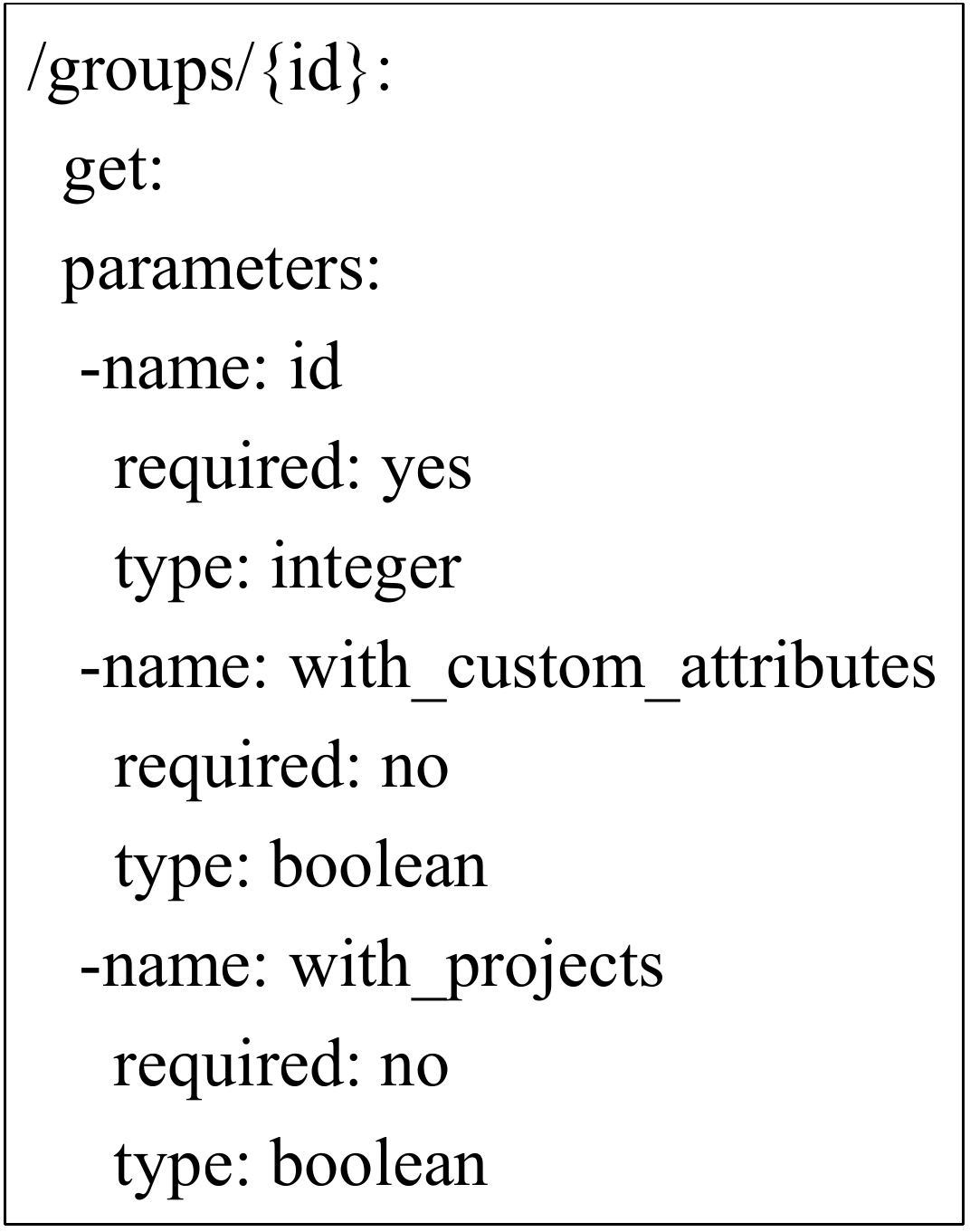}
         \caption{The corresponding Swagger specification.}
         \label{fig:specflow2}
     \end{subfigure}
     \hfill
     \begin{subfigure}[b]{0.20\textwidth}    
         \centering
         \includegraphics[width=1.28in,height=1.38in]{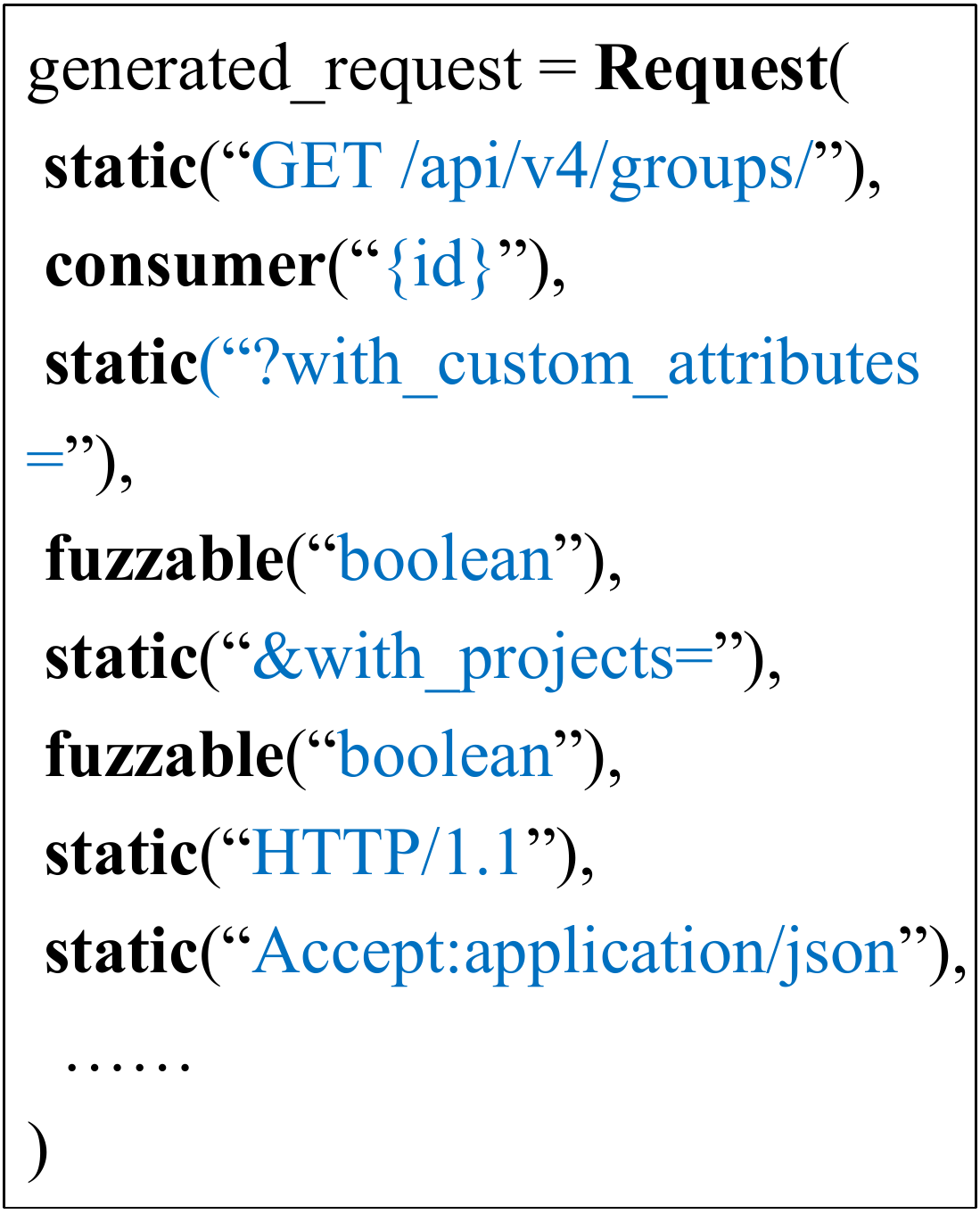}
         \caption{The request template for "GET /groups/\{id\}".}
         \label{fig:specflow22}
     \end{subfigure}
     \hfill
     \begin{subfigure}[b]{0.18\textwidth}    
         \centering
         \includegraphics[width=1.23in,height=1.38in]{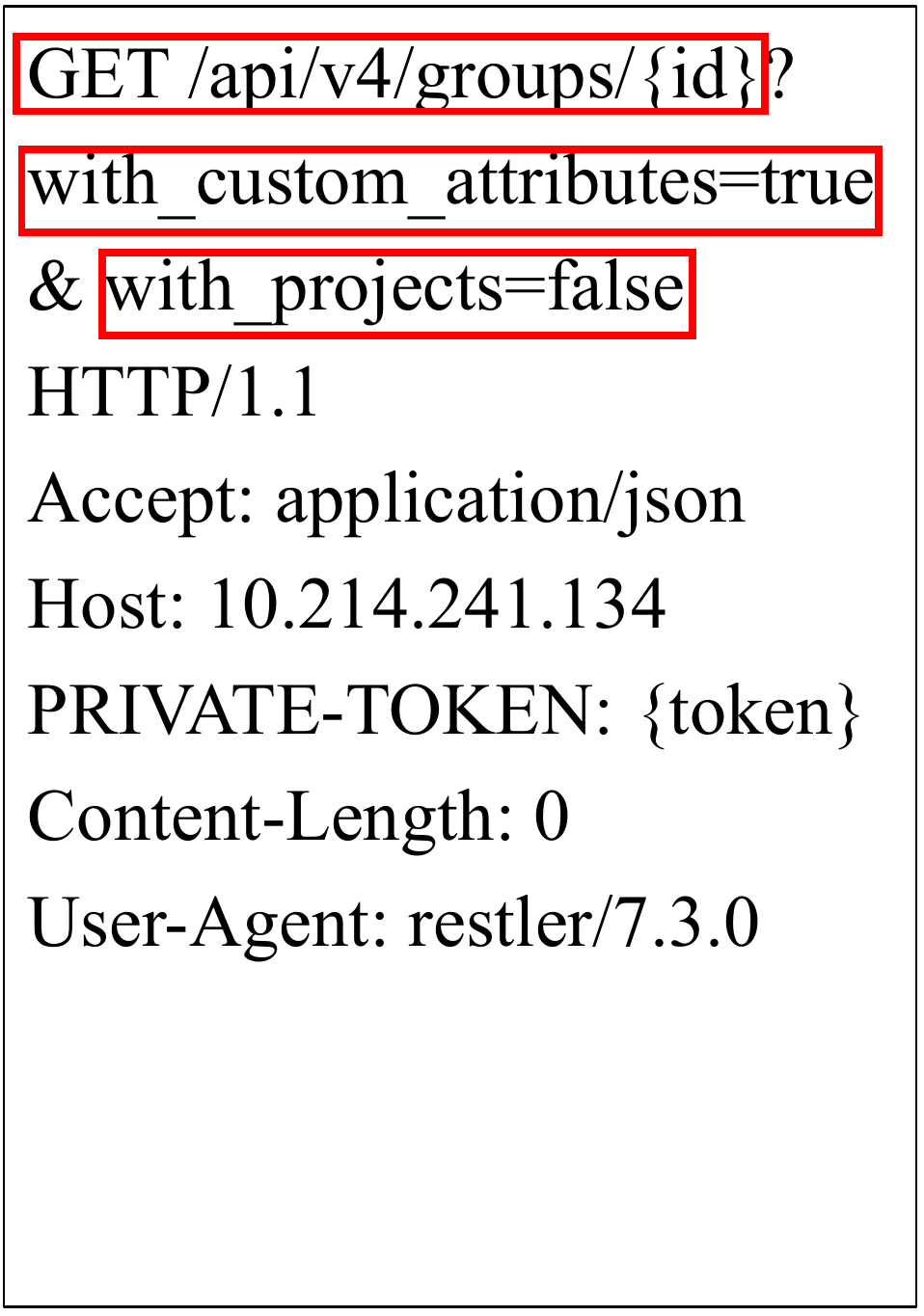}
         \caption{The  generated request example.}
         \label{fig:specflow3}
     \end{subfigure}
     \hfill
        \begin{subfigure}[b]{0.18\textwidth}    
            \centering
            \includegraphics[width=1.23in,height=1.38in]{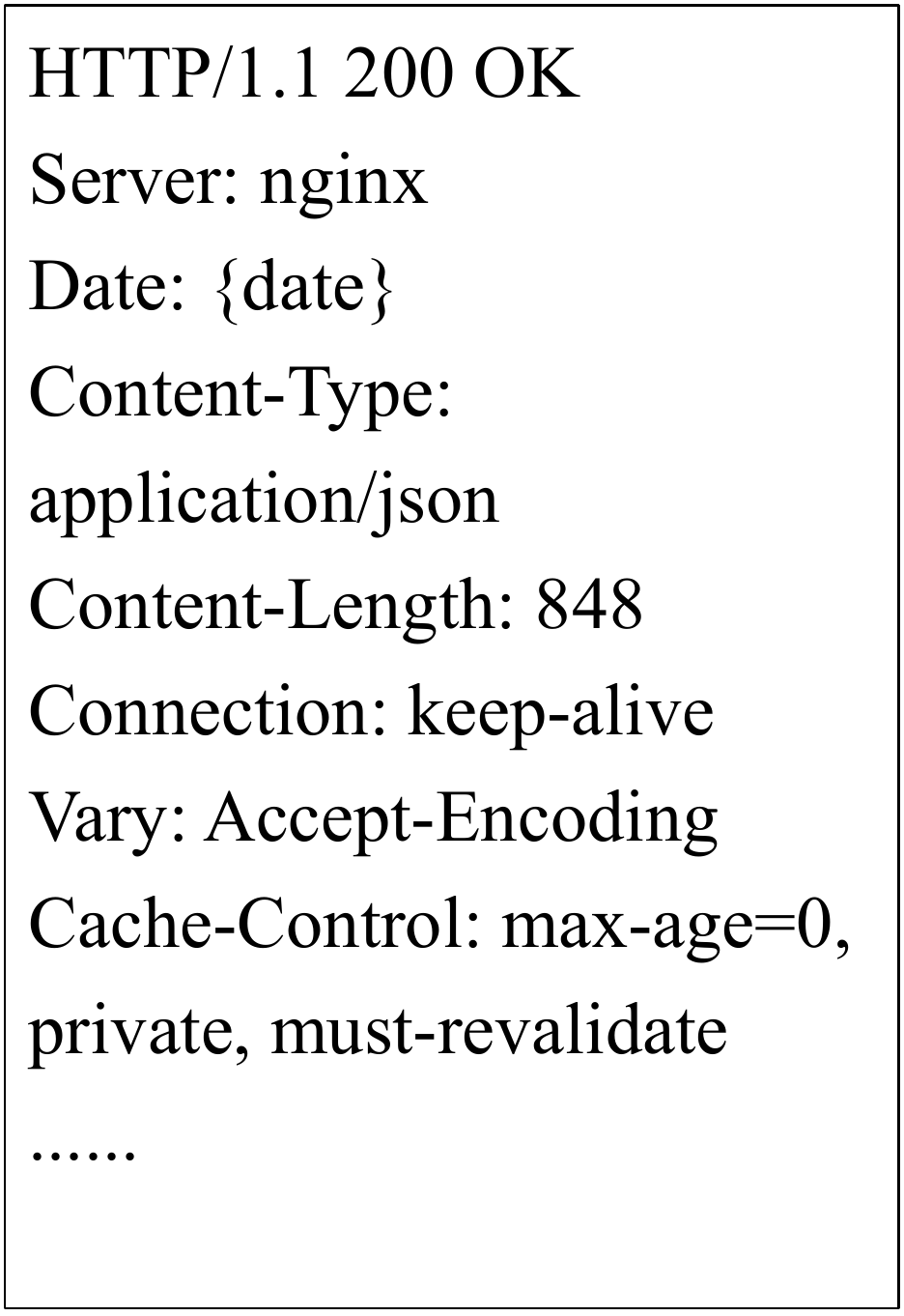}
            \caption{The corresponding response.}
            \label{fig:specflow4}
        \end{subfigure}
        \caption{An example to convert the description of a request on the web into a request template, and what a generated request and its corresponding response look like.}\label{fig:specflow}
\end{figure*}

\noindent\textbf{\texttt{Compiler} \texttt{Module:}} 
A REST API fuzzer implements the \texttt{Compiler Module} to generate a fuzzing grammar and construct request templates, 
whose procedures are as follows. 
First, the fuzzer requires the Swagger specification for each request type. 
As shown in Fig.~\ref{fig:specflow}, to obtain the Swagger specification, 
a user 1) can read the description for each request type published on the web page by vendors, 
and  2) manually transforms the description into the Swagger specification 
as shown in Fig.~\ref{fig:specflow1} and Fig.~\ref{fig:specflow2}.  
Then, based on the Swagger specification, 
the fuzzer performs a lightweight static analysis to construct \emph{request templates} as shown in Fig.~\ref{fig:specflow22}. 
To be specific, it 1) infers the dependencies among different request types; 
2) constructs a template for each request type; 
and 3) manually constructs a dictionary of alternative values for each parameter in a request template.  
For instance, as shown in Fig.~\ref{fig:specflow2}, 
a user can provide \{"\textit{\textbf{3}}", "\textit{\textbf{true}}", "\textit{\textbf{false}}"\} as the dictionary for the parameter "\textit{\textbf{with\_projects}}", 
where "\textit{\textbf{3}}" is the default value that does not match the parameter type. 

In particular, some request templates require target object ids of target resources, e.g., 
the "\{\textbf{\textit{id}}\}" as shown in Fig.~\ref{fig:specflow3}. 
If we provide static alternative values for these object ids, the fuzzer may suffer from the following two situations. 
1) The fuzzer can only test the same resources on a cloud service, which limits the exploration space; 
2) The resources can be deleted by \texttt{DELETE} requests, which hinders other requests for access to these resources. 
Therefore, the REST API fuzzer 
solves the problem by reading the target object id from the response of the previous request in a sequence. 
Specifically, it constructs special request templates to produce and consume target object ids. 
To produce a target object id, the fuzzer analyzes the response of a request like \texttt{GET} and \texttt{POST}, 
and stores the target id in memory, which will be freed when the test of the current request sequence ends. 
Then, if a latter request template needs a target object id, the fuzzer generates it by reading the stored id from memory. 
In general, a request that provides a target object id is named \emph{producer},  and a request that requires a target object id is named \emph{consumer}.

\noindent\textbf{\texttt{Sequence} \texttt{Module:}} 
A fuzzer combines different requests to construct different test sequences, e.g., it can send a sequence containing \texttt{POST} and \texttt{PUT} requests to create a resource and then modify this resource. 
To achieve this, it implements the \texttt{Sequence Module}  to dynamically extend request sequence templates started from an empty one, whose details are as follows. 
1) When entering the \texttt{Sequence Module}, a fuzzer starts with an empty sequence template;  
2) It  maintains a set of successfully extended sequence templates, and  performs the extension process to construct new sequence templates based on these successfully extended templates.  Specifically,  it adds each candidate request at the end of the current sequence template and constructs new candidate sequence templates.  
Then, the fuzzer empties the set and waits for the new successfully extended sequence templates;  
3) For each candidate sequence template, the fuzzer utilizes the \texttt{Generation Module},  which will be introduced later,  to generate each request in the sequence template  and then tests a target cloud service;  4) It analyzes the response of each request to infer the triggered behavior. 
For instance, a response of \texttt{400} Bad Request, \texttt{200} OK or \texttt{500} Internal Error represents that the cloud service performs the following behavior, respectively: 
The cloud service refuses to execute the request which does not pass the checking, the cloud service executes the request normally, or 
it executes the request abnormally which results in an unexpected error; 
5) Generally, the fuzzer checks the response of the last request in each candidate sequence to infer whether the sequence is valid. 
If the last request 
gets a response in \texttt{20$\times$} Range from the cloud service, the fuzzer considers the candidate sequence successfully extended and adds it to the set of successfully extended sequence templates;  
6) If the set is not empty, the fuzzer goes to step 2) to perform the next extension process based on the newly collected sequence templates. Otherwise, it goes to step 1) to restart the extension process with an empty sequence template. Note that in the \texttt{Sequence Module}, the constructed sequence templates always satisfy the dependencies between producers and consumers.

\noindent\textbf{\texttt{Generation} \texttt{Module:}} 
A REST API fuzzer implements this module to generate each request in a sequence template. 
Specifically, the fuzzer assigns a value for each parameter in the request, 
and constructs a complete request that is ready to be sent to a cloud service. 
There are two methods to obtain a parameter value, 
which are 1) selecting an alternative value from the dictionary
or 2) reading a target object id from a response of a previous request. 
The following instance shows how the fuzzer obtains parameter values using the above two methods. 
As shown in Fig.~\ref{fig:specflow}, the request template \texttt{GET /groups/\{id\}} contains 3 parameters needed to be set. 
To generate a ready-to-use request, the fuzzer selects "\textit{\textbf{true}}" and "\textit{\textbf{false}}" as the values of "\textit{\textbf{with\_custom\_attributes}}" and "\textit{\textbf{with\_projects}}", respectively.  
As for the parameter "\textit{\textbf{id}}", the fuzzer will read a target object id from the response of a producer like \texttt{POST}, which provides an object id after creation.

 \noindent\textbf{\texttt{Checker} \texttt{Module:}} 
To improve error discovery of REST API fuzzing, Atlidakis et al. introduced several rule checkers to capture specific security rule violations~\cite{atlidakis2020checking}, which are implemented as the \texttt{Checker Module} in the REST API fuzzer. For instance, a fuzzer implements a use-after-free rule checker to find the request sequences triggering use-after-free bugs,  
 which access deleted resources in the cloud service scenario and can lead to private data leakage. 
An example process is as follows. The rule checker first sends a \texttt{POST} request to create a resource on a cloud service. Then, it sends a \texttt{DELETE} request to delete this resource. Finally, the fuzzer sends an access request, like \texttt{GET}, to access this deleted resource. 
If the cloud service 
does not return a response in \texttt{40$\times$} Range, 
the fuzzer considers that the use-after-free violation is triggered and saves the used request sequence locally for future analysis. 
Therefore, a REST API fuzzer tries to trigger specific rule violations in the \texttt{Checker Module}.


By utilizing the above main modules, existing REST API fuzzers automatically generate request sequences to test a cloud service via its REST APIs, and record the request sequences locally that trigger unique errors hidden in different states. However, they still have limitations in sequence extension and request generation, 
resulting in slow state exploration progress. 
On the contrary, we observe that multiple kinds of historical data collected during the fuzzing process are valuable to guide REST API fuzzing and solve the above limitations, which is the main focus of this paper.

\subsection{Neural Network Model}

As marked by the red boxes in Fig.~\ref{fig:specflow3}, 
a request consists of an endpoint/name, 
several parameters and their values. 
Intuitively, we can regard these elements in the red boxes as \textit{words} and transform a request into a word sequence. Then, a request generation can be transformed into a sequence generation, which is one of the common applications for neural network models~\cite{hochreiter1997long, devlin2018bert}. 
Specifically, a neural network model leverages neurons to construct complex computational networks~\cite{cho2014learning, chung2014empirical}. 
During training, the model adjusts the weights between neurons to fit the implied relationship between words in each sequence. 
Given an input word, the model improves the prediction probability of the ideal output word.
After training, the model predicts words one by one, which follows the implied relationship learned from the training set. 

In our application, we collect the parameters and their values used in the valid requests, which pass the checking of a cloud service, as the key mutations. 
Then, we leverage a neural network model to learn the implied relationship between these key mutations, i.e., 
which key mutations should be used together on a request. 
After training, the model predicts the key parameters to be mutated and their appropriate values for request generation. 
More details are shown in Section~\ref{sec:model}.

\section{Design of \sysname}\label{sec:design}

\subsection{Why Valid Long Sequence Matters}\label{sec:motivation}

As previously mentioned, different requests can be sent in a sequence to a cloud service to trigger a series  of behaviors and further explore unique errors hidden in deep states. 
For instance, in a sequence, \texttt{POST}, \texttt{PUT}, and \texttt{DELETE} requests can be sent orderly via GitLab Projects API. 
Then, GitLab will create a new project, modify this project's parameters, and delete this project, respectively. 
We can conclude from the above instance that 
a long sequence covers the possible request combinations used in a short sequence, as does the range of states that can be reached. 
Therefore, to explore the errors hidden in hard-to-reach states, an ideal REST API fuzzer is expected to send long sequences more times, and should not always test a cloud service using short sequences. 

\begin{figure}
  \setlength{\abovecaptionskip}{0.02cm}
  \setlength{\belowcaptionskip}{-0.7cm}
\centering
\includegraphics[width=2.4in,height=1.5in]{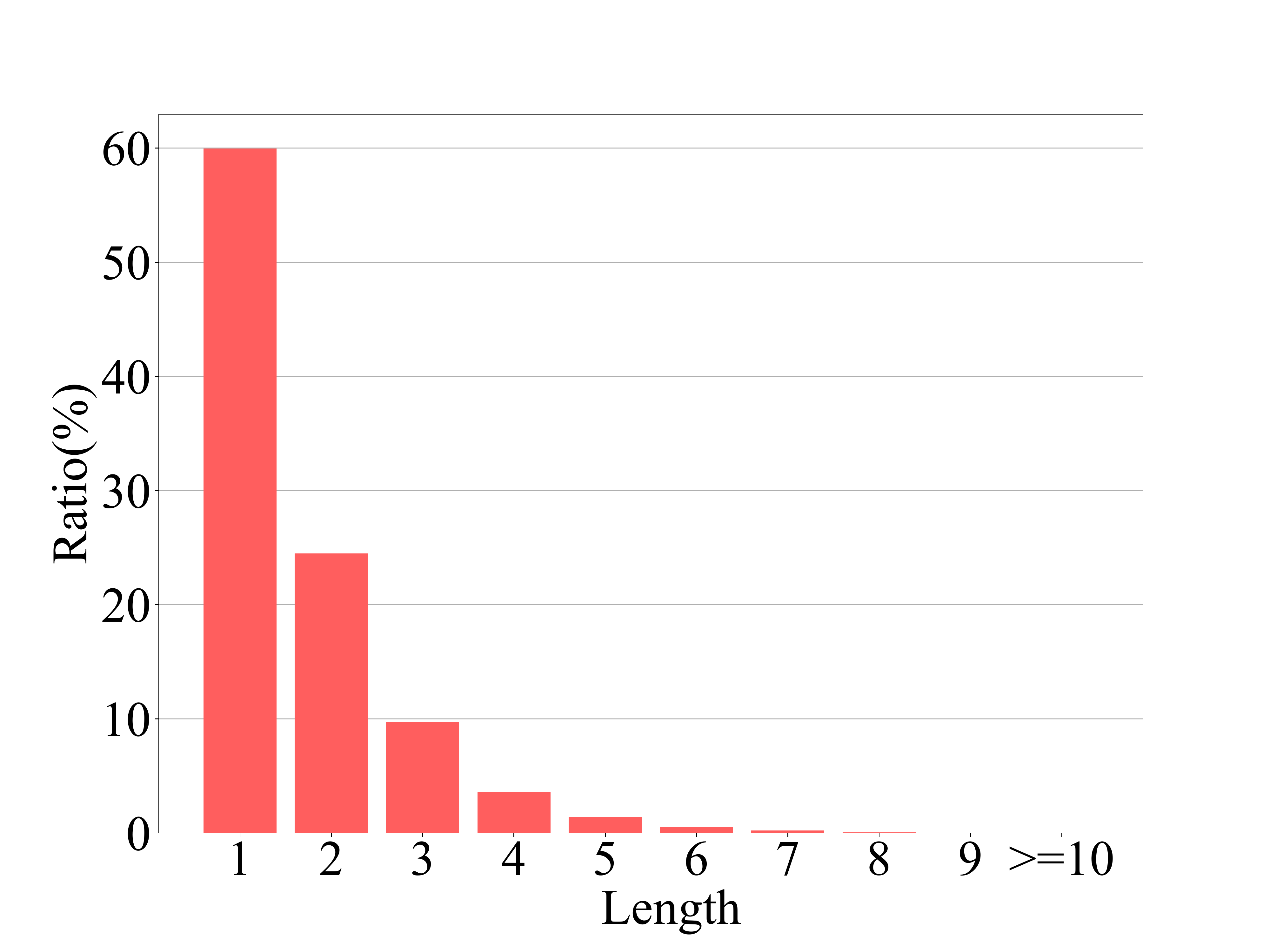}
\caption{The distribution of request sequence length sent by RESTler when fuzzing GitLab Projects API.}
\label{fig:length_dist}
\end{figure}

To further motivate our research, 
we 
analyze the REST API-related issues on GitLab. 
To be specific, 
we collect the error issues related to the available resources for GitLab REST APIs, 
and count the number of issues 
triggered by a request sequence or a singular request, respectively. 
Then, we realize that the issues caused by a request sequence account for the majority.   
In all 56 issues, the proportion of the issues caused by request sequences is 76.79\%. 
For instance, 
a published error 
is triggered by the following steps. 
1) Create a GitLab project; 
2) Delete this project; 
3) Access this deleted project with a \texttt{GET} request. 
Since the last request achieves to access a deleted resource, 
this error is considered 
a use-after-free bug in the cloud service scenario 
and is confirmed by the vendors of GitLab. 
It takes at least 3 requests in a sequence to reproduce this published error. 
The case study demonstrates that a REST API fuzzer is expected to 
test a target cloud service with long request sequences. 
However, we find that  
existing REST API fuzzers cannot effectively generate long sequences to test a cloud service. 
To illustrate this, we conduct the following case study to show the proportion of 
request sequence length 
executed in the fuzzing process. 
We utilize \textit{RESTler}~\cite{restler}, 
which is one of the most famous REST API fuzzers and implements the main modules as shown in Section~\ref{sec:restapi},  to fuzz GitLab Projects API. 
Then, we construct the length distribution of the generated request sequences. 
The evaluation lasts for 12 hours, 
and the proportion of each sequence length is shown in Fig.~\ref{fig:length_dist}, 
from which we can learn that the length of most request sequences generated by RESTler is less than 3. 
The proportion of the sequences with lengths of 1 and 2 is around 85\%. 

These results demonstrate that existing REST API fuzzers hardly generate long sequences to trigger series of behaviors on cloud services. 
Especially, it is hard for them to trigger security bugs explored by long request sequences as shown in~\cite{atlidakis2020checking}. 
We infer the reasons as follows. 
First, existing fuzzers cannot effectively generate {valid} requests that can pass a cloud service's syntax and semantic checking. 
For each parameter in a request template, 
they cannot figure out the appropriate parameter value, 
but randomly select an alternative value from a preset dictionary, e.g., 
\{"\textit{\textbf{3}}", "\textit{\textbf{true}}", "\textit{\textbf{false}}"\} for  "\textit{\textbf{with\_projects}}" as exemplified  in  Section~\ref{sec:restapi}.  
Thus, 
their generated requests are more likely to receive responses in \texttt{40$\times$} Range from the cloud service.  
This fact is also confirmed by the results in Section~\ref{sec:fuzzingperformance}, which demonstrate the low pass rate of the requests generated by existing REST API fuzzers. 
Then, due to the lack of a proper mechanism to generate valid requests, existing fuzzers are difficult to extend a sequence template to form a long one. 
As shown in Section~\ref{sec:restapi}, the fuzzers frequently restart the sequence extension process with an empty sequence template, if the last request in a sequence fails to pass the checking. 
Moreover, the kinds of request combinations in a short sequence template are limited, which can be fully explored by existing fuzzers in the fuzzing process. 
Therefore, most of the time, 
existing REST API fuzzers tend to test a cloud service 
with short and already-used request sequence templates.

\noindent\textbf{Motivation.}
Based on the above analysis, 
we discover that most published errors 
 are triggered by request sequences. 
While existing REST API fuzzers are hard to generate long sequences with acceptable request generation quality. 
Most of the time, 
existing fuzzers test target cloud services with repetitive short request sequences, 
limiting their performance of discovering unique errors. 
In particular, it is hard for them to explore security bugs  triggered by long request sequences~\cite{atlidakis2020checking}. 
Therefore, to improve security bug discovery 
 on cloud services, 
 a new fuzzing solution is demanded to 
 improve the sequence construction and request generation. 

\subsection{Framework of \sysname}\label{sec:framework}

Motivated by the above findings, 
we propose a hybrid data-driven solution named \textit{\sysname} as follows. 
1) \sysname implements the \emph{length-orientated sequence construction}, 
and leverages the used request sequences as  seed sequence templates 
to guide sequence construction; 
2) \sysname implements the \emph{attention model-based request generation}, 
and utilizes an attention model to provide proper values for the key parameters, 
which increases the  pass rate to the cloud service's checking; 
3) To find errors caused by incorrect parameter usage, 
\sysname implements the \emph{request parameter violation checking}, 
and constructs a new security rule checker named \emph{DataDriven Checker} to 
generate requests with undefined parameters. 
The above three designs work together to improve the fuzzer's performance on cloud services.

 \begin{figure*}[t] 
  \setlength{\abovecaptionskip}{0.02cm}
 \setlength{\belowcaptionskip}{-0.8cm}
\begin{center} 

\includegraphics[width=6.1in,height=1.3in]{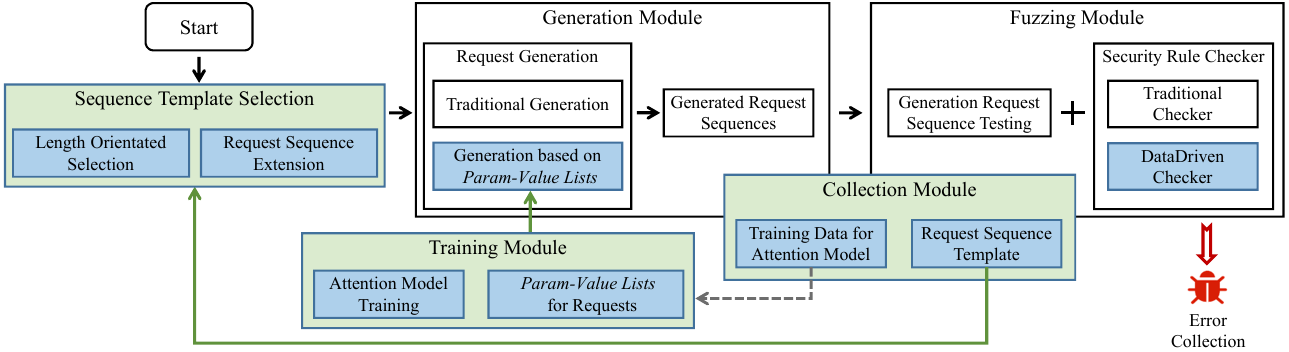}

\caption{The framework of \sysname.}\label{fig:framework}

\end{center} 
\end{figure*}

As shown in Fig.~\ref{fig:framework}, the framework of \sysname contains 5 main components: 
\texttt{Sequence} \texttt{Template} \texttt{Selection}, 
\texttt{Generation} \texttt{Module}, \texttt{Fuzzing} \texttt{Module}, 
\texttt{Collection} \texttt{Module}, and \texttt{Training} \texttt{Module}. 
The basic workflow  is that \sysname continuously generates request sequences 
to test a  cloud service via its REST API. 
Before fuzzing, like existing fuzzers, 
\sysname constructs request templates for a target REST API with the \texttt{Compiler Module} 
as described in Section~\ref{sec:restapi}. 
For each test, 
\sysname 1) utilizes the \texttt{Sequence} \texttt{Template} \texttt{Selection} to probabilistically select request sequence templates according to their lengths
and then extend sequences; 
2) uses the \texttt{Generation} \texttt{Module} to generate each request in sequence templates and constructs ready-to-use request sequences; 
and 3) utilizes the \texttt{Fuzzing} \texttt{Module} to test a cloud service with generated request sequences. 
During the execution of the above modules, 
the \texttt{Collection} \texttt{Module} collects the following historical data: 
First, \sysname collects the {valid} request sequences, all of whose requests pass the cloud service's checking, to enrich the sequence template set. 
Second, if a request passes the checking, 
\sysname collects the mutated parameter values that are not the default ones.
For clarity, we use "\textit{param-value pair}" to represent the mutated parameter and its used value in request generation, e.g., <"\textit{\textbf{with\_projects}}", "\textit{\textbf{false}}"> as shown in Fig.~\ref{fig:specflow}. 
Then, \sysname periodically invokes the \texttt{Training} \texttt{Module} to train an attention model with the collected \textit{param-value pairs}. 
After training, \sysname uses the model to generate {lists of desirable \textit{param-value pairs}} 
for each request template to enhance the request generation. 
In the following, we use "\emph{param-value list}" to represent 
a list of generated \textit{param-value pairs}. 
The details of each module are  as follows. 

\noindent\textbf{\texttt{Sequence} \texttt{Template} \texttt{Selection:}} 
In this component, \sysname constructs candidate sequence templates 
based on the seed sequence templates collected from the \texttt{Collection Module}. 
Specifically, \sysname first assigns a different selection probability 
to each seed sequence template according to its length. 
The longer a sequence template is, the larger its selection probability is. 
Thus, \sysname obtains a set of selected sequence templates that tend to have larger lengths. 
Second, for each selected sequence template, 
\sysname performs the extension process to construct a candidate sequence template. 
In other words, 
\sysname adds a new request template 
at the end of the selected template. 
Based on the above design, \sysname implements the \emph{length-orientated sequence construction}. 
In implementation, we use $log_{10}(l + 1)$, where $l$ is the length of a sequence template, 
as the probability function for the \texttt{Sequence Template Selection}. 

After finishing all the extension processes, 
\sysname obtains a set of candidate sequence templates and enters the \texttt{Generation Module}.

\noindent\textbf{\texttt{Generation} \texttt{Module:}} 
In this component, 
\sysname generates each request in a candidate sequence template, i.e., 
\sysname assigns a value for each parameter excluding a target object id in a request. 
As shown in Fig.~\ref{fig:framework}, 
\sysname implements two methods to generate requests. 
The first one is the traditional generation method as described in Section~\ref{sec:restapi}, 
i.e., randomly selecting an alternative value for each parameter. 
On the other hand, the procedure of the second method is as follows. 
To generate a request, 
\sysname 1) selects the default value for each parameter; 
2) finds the set of \textit{param-value lists}, 
each of which contains a list of \textit{param-value pairs} for mutation, 
generated by the attention model for this request template; 
3) probabilistically selects a candidate \textit{param-value  list} from the set according to a uniform distribution; 
and 4) 
uses the specified values to modify the parameters 
according to the \textit{param-value pairs} in the selected list. 
For the parameters that are not in the selected \textit{param-value list},
\sysname remains their default values in request generation. 
By performing the above procedure, \sysname achieves the \emph{attention model-based request generation}.  

In implementation, 
to generate a sequence containing $n$ requests, 
\sysname uses \emph{param-value lists} to generate the first $n - 1$ requests, 
and uses the traditional generation method to generate the last request. 
The reasons are as follows. 
When generating the first $n - 1$ requests, 
\sysname would like to leverage \emph{param-value lists} to improve the generation quality. 
This helps trigger normal executions of a target cloud service, 
and obtain a target object id if there is a producer in the request sequence. 
Then, \sysname utilizes the traditional generation method to generate the last request, 
in order to increase the probability of triggering an abnormal behavior, e.g., tries to create a resource with the parameter value that does not conform to the defined type. 

After generating all the requests, 
\sysname obtains ready-to-use request sequences, and enters the \texttt{Fuzzing Module}.

\noindent\textbf{\texttt{Fuzzing} \texttt{Module:}} 
In this component, \sysname performs the following process to explore unique errors. 
First, \sysname sends a generated request sequence to a cloud service via its REST API, 
and analyzes the corresponding response to each request. 
If \sysname receives a response in \texttt{50$\times$} Range, 
\sysname thinks that the corresponding request 
triggers an error, and 
stores the generated sequence among with their responses locally for future analysis. 
Second, \sysname leverages each security rule checker to mutate the current request sequence, 
in order to trigger specific rule violations. 
In particular, 
we propose the \emph{request parameter violation checking}, 
and implement a new security rule checker named \emph{DataDriven Checker}  
to find the errors caused by incorrect parameter usage, 
whose target rule violation is as follows. 

In general, vendors define specific parameters for each request 
 to trigger specific behaviors of a cloud service. 
If a user adds an undefined parameter to a request and sends it to a cloud service, 
generally the cloud service ignores the incorrect usage of the undefined parameter, 
and 
performs  behaviors according to the values of the defined parameters. 
However, a cloud service may perform unexpected behaviors according to the undefined parameters due to unexpected conditions, e.g., parameter definition update and incorrect code implementation. 
The unexpected behaviors can trigger security-related  exceptions, 
e.g., a cloud service may access a non-existent resource. 
As a result, the process of a cloud service can crash and return a response in \texttt{50$\times$} Range due to undefined parameters. 

To capture this kind of rule violation, 
the \emph{DataDriven Checker} adds an undefined \textit{param-value pair}
to a request during the generation process. 
Specifically, 
for the last request in a sequence, 
\sysname randomly selects an undefined \textit{param-value pair} collected
in the \texttt{Collection} \texttt{Module}. 
Then, \sysname adds the selected  pair into the last request of the sequence, e.g., adds <"\textbf{\textit{min\_access\_level}}", "\textbf{\textit{1}}"> in the request shown in Fig.~\ref{fig:specflow3}. 
Finally, \sysname sends the sequence with the newly constructed request 
to a cloud service via its REST API. 
If the cloud service returns a response in \texttt{50$\times$} Range, 
\sysname considers that the new request 
triggers an incorrect parameter usage error, 
and stores the sequence locally for further analysis. 

After executing the above process for all the generated request sequences, 
\sysname enters the \texttt{Sequence Template Selection} to start the next iteration.

\noindent\textbf{\texttt{Collection} \texttt{Module:}} 
In the fuzzing process, 
\sysname first collects the valid 
request sequences, 
in which  all the requests pass the cloud service's checking in the \texttt{Fuzzing Module}. 
These sequences are used as the seed sequence templates 
in the \texttt{Sequence Template Selection} to guide the sequence construction. 
Second, \sysname analyzes the response of each generated request from a cloud service, 
and extracts the \textit{param-value pairs} used in the valid requests that pass the checking. 
Specifically, 
if a request gets a response in \texttt{20$\times$} and \texttt{50$\times$} Range, which means that the request passes the checking and triggers a behavior of the cloud service, 
\sysname analyzes the used value for each parameter in the request. 
If the used value is not the default one for a parameter, 
\sysname regards the value as a key mutation on this parameter, 
and stores them as a \textit{param-value pair}. 
Thus, \sysname collects a list of \textit{param-value pairs} for each valid request, 
which contains the valid mutations to help pass the cloud service's checking. 
For instance, \sysname will collect [<"\textit{\textbf{with\_custom\_attributes}}", "\textit{\textbf{true}}">, <"\textit{\textbf{with\_projects}}", "\textit{\textbf{false}}">] for the example in Fig.~\ref{fig:specflow3}. 
The collected lists are used to 1) train the attention model in the \texttt{Training} \texttt{Module}, 
and 2) provide undefined \textit{param-value pairs} for the \emph{DataDriven} \emph{Checker} 
to trigger incorrect parameter usage errors.

\noindent\textbf{\texttt{Training} \texttt{Module:}} 
\sysname periodically invokes the \texttt{Training} \texttt{Module} to train an attention model 
with the \emph{param-value pairs} collected  
in the \texttt{Collection} \texttt{Module}. 
After training, \sysname utilizes the model to generate the \textit{param-value lists} for each request template, 
which will be used in the \texttt{Generation} \texttt{Module} to generate requests. 
In the following, we describe the detailed usage of the attention model. 


\subsection{Attention Model used in \sysname}\label{sec:model}

In this paper, we utilize a customized attention model, 
 which is one of the most powerful machine learning models for sequence generation,  
to learn the implicit relationship between \textit{param-value pairs}. 
Then, we utilize the model to explore more ideal combinations of \textit{param-value pairs}, which are denoted as \textit{param-value lists}. 
To achieve this, 
in the \texttt{Training} \texttt{Module}, 
\sysname trains the attention model 
using the lists of \textit{param-value pairs} as the training data, 
which are collected from the requests triggering responses in \texttt{20$\times$} Range. 
The process is as follows. 

\begin{figure}[t] 
\setlength{\abovecaptionskip}{0.1cm}
 \setlength{\belowcaptionskip}{-0.7cm}
 \centering
\begin{center} 

\includegraphics[width=3.2in,height=1.5in]{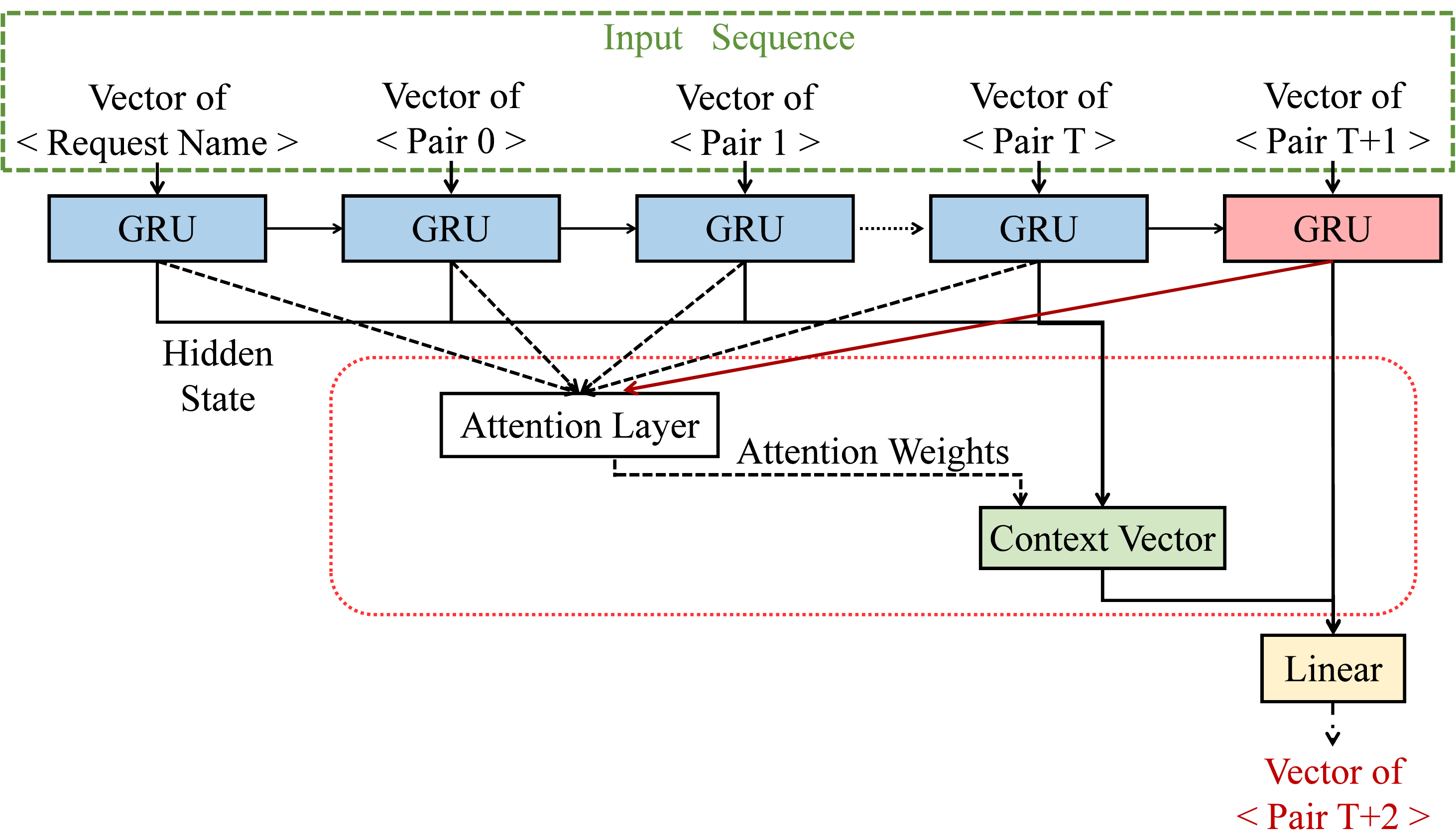}

\caption{The structure of the 
attention model used in \sysname.}\label{fig:structure}

\end{center} 
\end{figure}

1) \sysname adds the corresponding request name 
in the front of each list of \textit{param-value pairs}, 
and regards each list as a word sequence, e.g., ["\texttt{GET /api/v4/groups/\{id\}}", "<\textit{\textbf{with\_custom\_attributes}}, \textit{\textbf{true}}>", "<\textit{\textbf{with\_projects}}, \textit{\textbf{false}}>"] for the example as shown in Fig.~\ref{fig:specflow}. 
Then, we follow the common processing of the machine learning model and embed all the words into vectors, e.g., "<\textit{\textbf{with\_projects}}, \textit{\textbf{false}}>" is represented by a unique vector in matrix form. 
Thus, 
the lists collected in the \texttt{Collection} \texttt{Module} 
are transformed into a set of vector sequences (similar to the input sequence 
as shown in Fig.~\ref{fig:structure}); 
2) \sysname randomly divides the vector sequences into the training set and the validation set, 
which are used to train the model and verify the prediction performance of the model, respectively; 
 3) In each set, 
\sysname utilizes a vector sequence to train/verify the attention model. 
Specifically, \sysname uses the previous $n$ vectors in the sequence as the input and the $n+1$ th vector as the expected output as shown in Fig.~\ref{fig:structure}. 
During the training process, 
the model adaptively adjusts the weights between neurons
and increases the prediction probability of the expected output vector. 
For instance, assuming the ideal output is <\textit{pair T+2}> as shown in Fig.~\ref{fig:structure}, the model will adjust the weights to improve the prediction probability of <\textit{pair T+2}> and lower others' prediction probabilities; 
 4) 
 After \sysname finishes a certain number of training epochs and improves the prediction accuracy to a stable level, the training process ends. 

After the training process, 
\sysname utilizes the attention model to generate the \textit{param-value lists} for the request templates, which provide the \textit{param-value pairs} as the training data. 
To be specific, 
\sysname employs 
the vector of a request name as the first input of the attention model, 
and predicts a \textit{param-value pair} as the output. 
Next, the predicted pair is 
appended to the previous input 
as the new input to predict the next pair. 
The end conditions for the generation of \textit{param-value pairs} are as follows. 
1) The attention model thinks that the sequence generation finishes and outputs a terminator 
(generally the terminator is a null vector that does not represent any word); 
2) The list of \textit{param-value pairs}  reaches the preset maximum length, which is significantly greater than the sequence length in the training set. 
Then, the generated list is stored locally, 
and will be used as a \emph{param-value list} in the \texttt{Generation Module}
as described in Section~\ref{sec:framework}. 

In implementation, 
we use a lightweight attention model, 
which contains a Gated Recurrent Unit (GRU) neural network~\cite{cho2014learning, chung2014empirical}, 
an attention layer~\cite{luong2015effective} 
and a linear layer as shown in Fig.~\ref{fig:structure}, to handle our generation problem. 
\sysname invokes the \texttt{Training Module} every two hours 
and retrains the model from scratch based on the newly discovered training data.  
The training and generation process of the attention model 
runs in parallel with the fuzzing process, 
which will not affect the fuzzing performance of \sysname. 
Contributed by the above settings, 
the time cost to train our model is less than 500 seconds as shown in Section~\ref{sec:overheadandpassrate}. 

\subsection{Data Flow of \sysname}

\begin{figure}[t] 
    \setlength{\abovecaptionskip}{0.02cm}
   \setlength{\belowcaptionskip}{-0.7cm}
  \begin{center} 
  
\includegraphics[width=3.5in,height=1in]{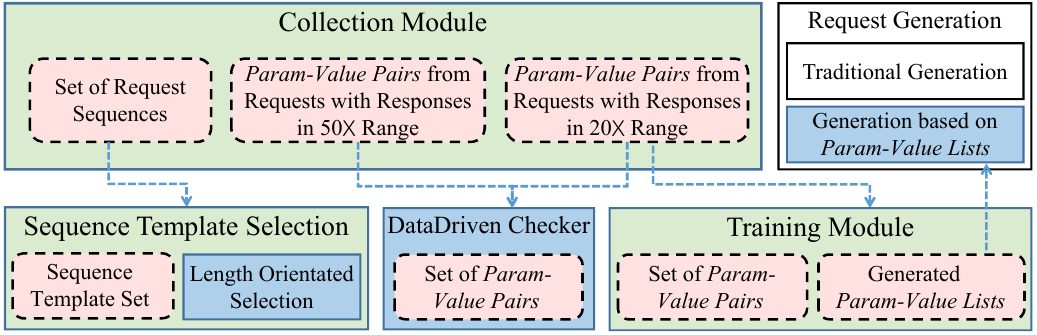}
  
  \caption{The data flow of the three new designs in \sysname.}\label{fig:dataflow}
  
  \end{center} 
  \end{figure}

In this subsection, we 
illustrate how the three core data-driven designs work together in \sysname from the perspective of data flow. 
As shown in Fig.~\ref{fig:dataflow},  in the \texttt{Collection Module}, 
\sysname mainly collects three datasets: 
1) the set of request sequences whose requests successfully pass the checking of a cloud service; 
2) the \textit{param-value pairs}  from the requests with responses in \texttt{50$\times$} Range;  
and 3) the \textit{param-value pairs}  from the requests with responses in \texttt{20$\times$} Range.  
In the following,  
we illustrate which design in \sysname utilizes these datasets. 

For the set of request sequences, \sysname uses it to guide sequence construction in the \texttt{Sequence} \texttt{Template} \texttt{Selection}. Since 
\sysname mainly deals with sequence template construction in this design, 
the other two datasets are not used.

The \textit{param-value pairs} from the valid requests, which pass the cloud service's checking and have responses in \texttt{50$\times$} and \texttt{20$\times$} Range, are used in the \emph{DataDriven Checker} to explore incorrect parameter usage errors. Since a request should pass the checking before triggering a potential error, in the \emph{DataDriven Checker} \sysname does not use the \textit{param-value pairs} from the requests having the responses in \texttt{40$\times$} Range, which may cause the generated request to be rejected by a cloud service and reduces the error discovery efficiency. 

The \textit{param-value pairs} from the requests, whose responses are in \texttt{20$\times$} Range, are also used in the \texttt{Training Module} as the training data of the attention model. Thus, the attention model tends to generate the \emph{param-value  lists}  that help pass the checking of a cloud service. Since 1) the requests having responses in \texttt{50$\times$} Range are pretty rare in the fuzzing process and 2) not all types of requests can get the responses in \texttt{50$\times$} Range, their number is not sufficient as the training data of a model. Thus, we do not include their \textit{param-value pairs} as the training data in our application scenario. On the contrary, if we collect sufficient requests with responses in \texttt{50$\times$} Range in the application scenarios like continuous fuzzing for a cloud service, we can utilize a machine learning model to learn how to generate a request triggering a response in \texttt{50$\times$} Range, and utilize the model to generate requests on demand.


Based on the above illustration from the perspective of data flow, we can learn that all the datasets for \sysname can be automatically collected during the fuzzing process.  
All the three designs rely on the datasets from the \texttt{Collection Module}, 
and are able to work together. 
Our hybrid data-driven approach is compatible with the fuzzing process, which can be deployed in general REST API fuzzers. 


\section{Evaluation}\label{sec:eval}

\subsection{Experiment Setup}\label{section:settings}

\begin{table}\scriptsize
  \setlength{\abovecaptionskip}{0.1cm}
\centering
\caption{The target cloud services and 11 REST APIs used in the experiments.}\label{table:target}
\begin{tabular}{m{0.9cm}<{\centering} m{0.7cm}<{\centering}  m{1.05cm}<{\centering} m{1.1cm}<{\centering} m{2.9cm}<{\centering}} 
\toprule
Cloud Service              & REST API        & \# Request Templates & \# Request Parameters & Description  \\ 
\toprule
\multirow{5}{*}{\begin{tabular}[c]{@{}c@{}}GitLab~\\14.1.0-ce.0.\end{tabular}} &Projects   &  29 & 304  &   API to interact with projects   \\
  &  Groups                           & 13        &  293      &    API related to groups            \\
&  Issues                             & 23        &  89  &     API to interact with    issues\\ 
& Commits                             & 14        & 59      &    { API related to commits }\\ 
 & Branches                           & 8       & 28    &    { API to interact with   branches}\\ 
\hline
\multirow{3}{*}{\begin{tabular}[c]{@{}c@{}}Bugzilla~\\5.0.4\end{tabular}}  & Comments       & 8      &   26           &   API to maintain comments           \\
&Bugs        & 11         &         89        &   API to maintain bug reports             \\
& Groups                                & 6           &     29           &   API to maintain user groups             \\
\hline
\multirow{3}{*}{\begin{tabular}[c]{@{}c@{}}WordPress~\\5.8.1\end{tabular}}& Categories       & 5     &            20       &    API to  maintain categories            \\
&Posts         & 10      &        42            &    API to interact with posts    \\
&  Comments                             & 8          &        40         &     API to maintain comments       \\
\toprule
\end{tabular}
\vspace{-2.0em}
\end{table}

In implementation, in order to measure the error discovery performance of our checker, 
we construct two prototypes without and with the \emph{DataDriven Checker}, 
which are named \sysnamepart and \sysname, respectively.

\noindent \textbf{Compared fuzzer.}
We evaluate \sysnamepart and \sysname against 
the state-of-the-art open-sourced fuzzer \emph{RESTler}~\cite{restler}, 
which is the first REST API fuzzer proposed in 2019 to explore errors on a cloud service automatically. 
Since another REST API fuzzer Pythia is not open-sourced, 
we cannot compare our fuzzers with it~\cite{atlidakis2020pythia}.

\noindent \textbf{Target cloud services.} 
We evaluate the above fuzzers on  GitLab~\cite{gitlab}, Bugzilla~\cite{bugzilla} and WordPress~\cite{wordpress} via 11 REST APIs as shown in Table~\ref{table:target}, and the reasons to select these targets 
are as follows. First, GitLab, Bugzilla and WordPress are widely-used cloud services. Evaluating their security is meaningful for vendors and users. 
Second, the request templates for the REST APIs of GitLab contain many parameters to support a wide variety of functionalities. 
On the contrary, the request templates of Bugzilla and WordPress are relatively simple ones with fewer parameters, which limits the place that can be mutated. 
Thus, in the evaluation, 
we can analyze the performance of fuzzers on the targets with different functional complexity. 
Third, each REST API  interacts with different kinds of resources on a cloud service. Fuzzing via these REST APIs, the evaluated fuzzers can execute different code lines and explore different states of a cloud service, which can examine their fuzzing performance more comprehensively. 

To construct the targets, 
we deploy an open-sourced version of each cloud service on our server, 
and construct request templates with the \texttt{Compiler Module}
for all the REST APIs, which is a one-shot effort and costs us about 1 hour.

\noindent \textbf{Experiment settings.} 
Each evaluation lasts for 48 hours on a docker container configured with 8 CPU cores, 20 GB RAM, 
Python 3.8.2, and the OS of 
Ubuntu 16.04 LTS. 
We run evaluations on 3 servers, 
each of which has two E5-2680 CPUs, 256GB RAM and a Nvidia GTX 1080 Ti graphics card.

\noindent \textbf{Evaluation metrics.} 
To measure the performance of each fuzzer on request generation and error discovery, 
we evaluate the  fuzzers with the following three metrics. 
First, to evaluate the request generation quality, we measure the pass rate of each fuzzer to the syntax and semantic checking of a cloud service, 
 which is calculated by dividing the number of the  responses in \texttt{20$\times$} and \texttt{50$\times$} Range by the number of total responses. 
Second, to measure how many unique request templates are successfully generated by each fuzzer, 
we count 
 the types of generated requests that 
 get responses in \texttt{20$\times$} Range. 
The more unique request templates are successfully generated and sent to a cloud service, 
the more kinds of behaviors are triggered by a REST API fuzzer. 
Third, we count the number of unique errors, 
which trigger the responses in \texttt{50$\times$} Range
or violate the defined security rules, 
reported by each fuzzer. 

\begin{table*}[t]\scriptsize
  \setlength{\abovecaptionskip}{0.1cm}
\centering
\caption{The fuzzing performance of RESTler, \sysnamepart and \sysname on 3 cloud services.}
\label{table:result}
\begin{tabular}{m{0.8cm}<{\centering} m{0.7cm}<{\centering} c m{0.6cm}<{\centering} c m{0.5cm}<{\centering}|m{0.6cm}<{\centering} c m{0.5cm}<{\centering}|m{0.6cm}<{\centering} c m{1.2cm}<{\centering}} 
\toprule
\multirow{2}{*}{\begin{tabular}[c]{@{}c@{}}Cloud\\ Service\end{tabular}}    & \multirow{2}{*}{\begin{tabular}[c]{@{}c@{}}REST\\ API\end{tabular}} & \multirow{2}{*}{\begin{tabular}[c]{@{}c@{}}\# Total Request\\ Templates~\end{tabular}} & \multicolumn{3}{c|}{RESTler} & \multicolumn{3}{c|}{\sysnamepart}  & \multicolumn{3}{c}{\sysname}  \\ 
\cline{4-12}
                           &                       &                                                                           & Pass Rate   & \begin{tabular}[c]{@{}c@{}}\# Unique \\Request Templates~\end{tabular} & Errors  & Pass Rate  & \begin{tabular}[c]{@{}c@{}}\# Unique \\Request Templates~\end{tabular} & Errors & Pass Rate   & \begin{tabular}[c]{@{}c@{}}\# Unique \\Request Templates~\end{tabular} & Errors$^{\rm a}$      \\ 
\toprule
\multirow{5}{*}{GitLab}    & Projects              & 29                                                                        & 72.01\%     & {22}                   & 7     &   95.77\%  & 26                &17     & 95.78\%  &   26       &    21 (9)      \\
                           & Groups                & 13                                                                        & 65.01\%     & {10}                   & 21   &   92.37\%  & 12                 &26     & 92.28\%  &    12      & 33 (10)         \\
                           & Issues                & 23                                                                        & 86.56\%     & {21}                  &   7   &   96.13\%  & 21                 &7     & 95.42\%  &    21      &  15 (6)        \\ 
                           & Commits               & 14                                                                        & 53.52\%     & 12                      &  16   &   86.65\%  & 12                 & 20    & 86.70\% & 12          & 37 (12)     \\
                           & Branches              & 8                                                                         & 81.10\%     & {8}                  & {2}&   {89.91\%}& {8}          & {3}    & {89.31\%}& {8}&  {9 (6)} \\
\hline
\multirow{3}{*}{Bugzilla} & Comments              &  8                                                                        &  88.79\%    &  7                   &   8   &  89.74\%   &  8                 &  8   &  90.23\%          &     8     &   8 (0)       \\ 
& Bugs                  &  11                                                                         & 45.03\%   &  4                   &  7    &    91.05\%  &  4               & 7    & 93.16\%          &      4    &  14 (6)     \\
                           & Groups                &  6                                                                         & 54.88\%    &  4                   &  5    &  74.01\%    &  5               &  6   & 72.85\%           &      5    &  11 (5)        \\
 \hline
\multirow{3}{*}{WordPress}    & Categories            &  5                                                                        &  75.33\%   & 4                    & 8      &  91.96\%   &  4      &   10  &    92.64\%        &   4       &   13 (4)      \\ 
& Posts                 &  10                                                                       &  94.06\%   &10                    & 4      &  95.13\%   &  10      &  5   &  95.56\%          &  10        &     7 (2)    \\
                           & Comments              &  8                                                                        &  96.61\%   &4                     & 0      &  99.65\%   &  4       &   0  &  99.43\%         &  4        &     0 (0)     \\ \hline
\multicolumn{3}{c}{Average}                                                                                                  & {73.90\%}   & {9.64} & {7.73} & {91.12\%} & {10.36} & {9.91} & {91.21\%} & {10.36} & {15.27} (5.45)\\
\toprule
\end{tabular}
\\\footnotesize{$^{\rm a}$The number of unique errors found by \sysname is presented in two parts: the total number of all the unique errors as shown in front of the parentheses, and the number of unique errors found by the \emph{DataDriven Checker} as shown in the parentheses.}\\
\vspace{-2.0em}
\end{table*}

\subsection{Fuzzing Performance Analysis}\label{sec:fuzzingperformance}

In this subsection, we analyze the fuzzing results reported by each fuzzer. 
The details are shown in Table~\ref{table:result}, 
from which we have the following conclusions.

$\bullet$
As shown in Table~\ref{table:result}, 
the new designs implemented in our fuzzers
significantly improve the pass rate on a cloud service.
For instance, the average pass rate of \sysnamepart and \sysname is 28.65\% and 28.28\% higher than RESTler on GitLab, respectively. 
The average pass rate of \sysnamepart is 35.03\% higher than RESTler on Bugzilla. 
Although the request templates of several APIs, e.g., WordPress Posts API, are easy-to-construct ones containing few parameters, 
our fuzzers still achieve higher pass rate compared to RESTler. 
The experimental results confirm the pass rate improvement contributed by our hybrid data-driven approach.

 $\bullet$
 \sysnamepart and \sysname can generate high-quality requests for more unique request templates 
than RESTler. 
For instance, the average number of unique request templates covered  by \sysnamepart is 8.22\% larger than  RESTler on GitLab. 
 The number of unique request templates covered by \sysname is 20.00\% larger than RESTler on GitLab Groups API. 
 We analyze the executions of each request template and have the following conclusions. 
 {Our hybrid data-driven approach significantly improves
 the complexity of generated request sequences.}  
 First, the \emph{length-orientated sequence construction} implemented 
 in \sysname increases the executions of long request sequences, 
 and increases the attempts on the consumers which require target object ids. 
Second, \sysname's \emph{attention model-based request generation} significantly improves request generation quality. 
 In particular, \sysname improves the producers' pass rate and obtains target object ids, 
 which can serve as the key parameter values for consumers. 
 As a result, \sysname 
 successfully sends more types of requests to a cloud service.  
 Since the only difference between \sysnamepart and \sysname is that \sysname 
 contains a new security rule checker \emph{DataDriven Checker}, 
 they perform the same on the number of covered request templates.

  $\bullet$
  Both \sysnamepart and \sysname find more unique errors than RESTler on most targets. 
For instance, \sysnamepart finds 37.74\% and 25.00\% more unique errors than RESTler on GitLab  and  WordPress, respectively. 
  We analyze the reasons as follows. 
  As indicated in Section~\ref{sec:motivation}, 
  the kinds of request combinations generated by RESTler are limited due to the short generated request sequences. 
  On the contrary, contributed by our approach, 
  \sysnamepart and \sysname can generate long sequences containing high-quality requests to test a cloud service. 
  The kinds of request combinations cover and far exceed those generated by RESTler. 
  Thus, 
  they can trigger more kinds of the cloud service's behaviors in a single request sequence test, 
  followed by finding more errors than RESTler. 
  In addition, 
  since \sysnamepart and \sysname can successfully generate requests for more unique request templates as shown in Table~\ref{table:result}, 
  they can trigger more and different behaviors of a target cloud service compared to RESTler. 
  Thus, they can trigger more unique errors based on these unique request templates. 
  Due to the randomness in fuzzing, 
  the number of unique errors found by \sysnamepart and \sysname is slightly different, 
  if we do not count the errors found by the \emph{DataDriven Checker} for \sysname.

\begin{table*}\scriptsize
  \setlength{\abovecaptionskip}{0.03cm}
 \centering
\caption{The real errors found by RESTler, \sysnamepart and \sysname on GitLab, Bugzilla and WordPress.}
 \label{table:realbug}
 \begin{tabular}{L{0.65cm} m{0.15cm}<{\centering} L{4.82cm} L{2.15cm} C{4.55cm} m{0.5cm}<{\centering} m{1.0cm}<{\centering} m{0.5cm}<{\centering}} 
 \toprule
 Target & No.~& \multicolumn{1}{c}{Endpoints of Used Requests}                                                                                                                                                     & \multicolumn{1}{c}{Responses}                                                                             &  Description                                                                                                                                                                                                                                                             & RESTler              & \sysnamepart                & \sysname          \\ 
 \toprule
\hspace{-0.1cm}\multirow{12}{*}[-30ex]{GitLab} & 1 & \begin{tabular}[c]{@{}l@{}}1. POST /api/v4/projects\\2.~GET /api/v4/projects/:id/fork\end{tabular}                                                                                                 & \begin{tabular}[c]{@{}l@{}}1. \texttt{201}~Created\\2.~\texttt{500} Internal Error\end{tabular}                             &  \begin{tabular}[c]{@{}c@{}} Use <"\emph{\textbf{namespace\_id}}", "\emph{\textbf{-1}}"> pair in the \\ second request. \end{tabular}                                                                                                                                                                                                              & \includegraphics[scale=0.01]{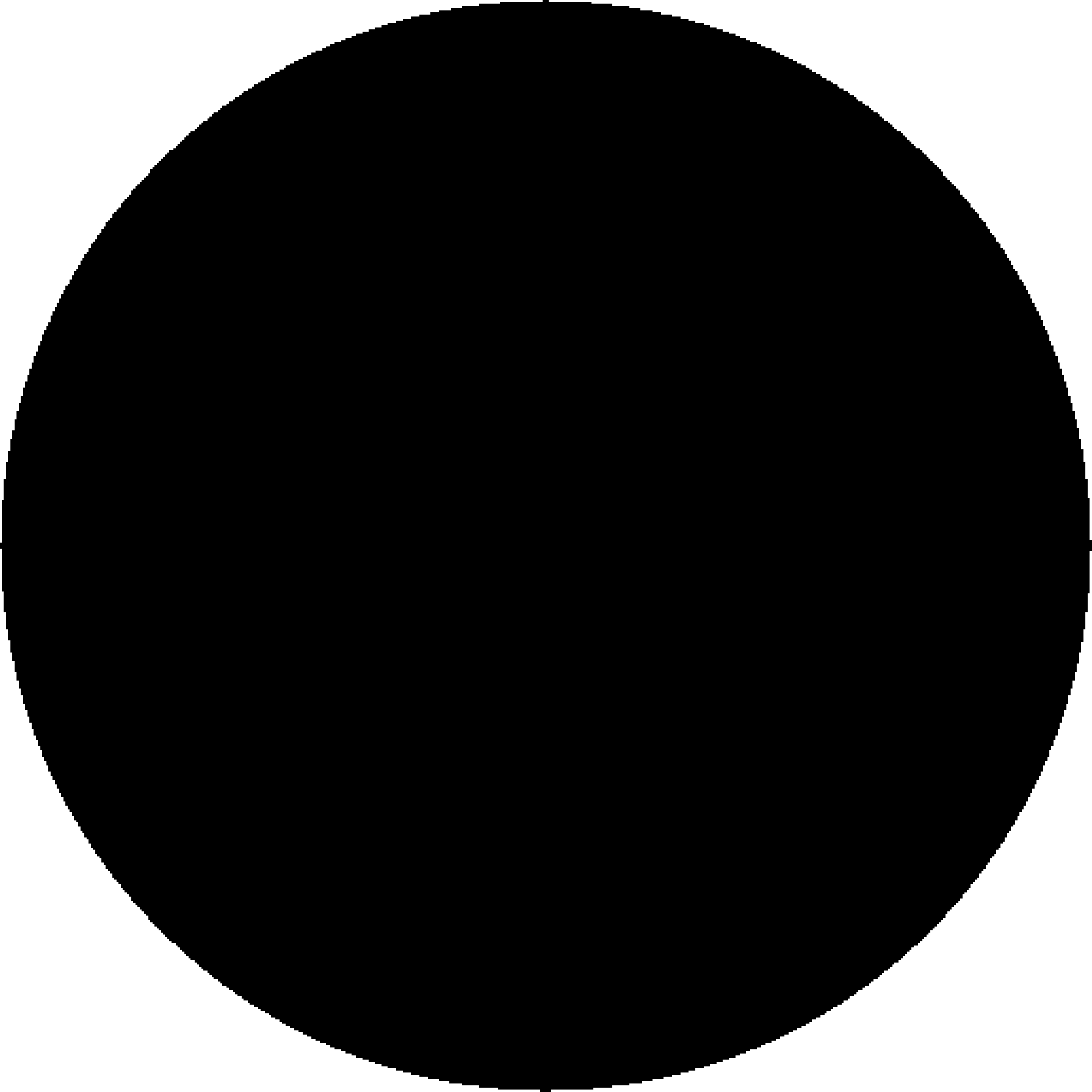}                 & \includegraphics[scale=0.01]{figure/mark1.pdf}                 & \includegraphics[scale=0.01]{figure/mark1.pdf}                  \\ 
 \cline{2-8} 
& 2 & \begin{tabular}[c]{@{}l@{}}1.~POST /api/v4/projects\\2.~PUT /api/v4/projects/:id\end{tabular}                                                                                                      & \begin{tabular}[c]{@{}l@{}}1. \texttt{201}~Created\\2.~\texttt{500} Internal Error\end{tabular}                             & \begin{tabular}[c]{@{}c@{}}The second request updates a project with the\\ undefined  parameter "\textbf{\textit{initialize\_}}\textbf{\textit{with\_readme}}".\end{tabular}  
&                      &                      & \includegraphics[scale=0.01]{figure/mark1.pdf}                  \\ 
 \cline{2-8} 
& 3  & \begin{tabular}[c]{@{}l@{}}1. POST /api/v4/projects\\2. GET /api/v4/projects/:id/custom\_attributes\\3. DELETE /api/v4/projects/:id\\4. GET /api/v4/projects/:id/custom\_attributes\\\end{tabular} & \begin{tabular}[c]{@{}l@{}}1.~\texttt{201} Created\\2. \texttt{200} OK\\3.~\texttt{202} Accepted\\4.~\texttt{500} Internal Error\end{tabular} & \begin{tabular}[c]{@{}c@{}} Use after free $^{\rm a}$. The fourth request tries \\to access the attributes of a deleted project.\end{tabular}                                                                                                                                                                                                      &                      & \includegraphics[scale=0.01]{figure/mark1.pdf}                 & \includegraphics[scale=0.01]{figure/mark1.pdf}                  \\ 
 \cline{2-8} 
& 4  & 1.~GET /api/v4/projects/:id/custom\_attributes                                                                                                                                                     & 1.~\texttt{500} Internal Error                                                                                     & Using string or~invlaid number as \textbf{\textit{:id}}.                                                                                                                                                                                                             & \includegraphics[scale=0.01]{figure/mark1.pdf}                 & \includegraphics[scale=0.01]{figure/mark1.pdf}                 & \includegraphics[scale=0.01]{figure/mark1.pdf}                  \\ 
 \cline{2-8} 
& 5  & \begin{tabular}[c]{@{}l@{}}1. POST /api/v4/projects\\2. POST /api/v4/projects/:proj\_id/hooks\\3. PUT /api/v4/projects/:proj\_id?/hooks/:hook\_id\end{tabular}                               & \begin{tabular}[c]{@{}l@{}}1. \texttt{201} Created\\2. \texttt{200} OK\\3.~\texttt{500} Internal Error\end{tabular}                  & \begin{tabular}[c]{@{}c@{}}The third request updates a hook with 3 \\  undefined  parameters "\textbf{\textit{requirements\_}}\\\textbf{\textit{access\_level}}", "\textbf{\textit{auto\_cancel\_pending\_}}\\\textbf{\textit{pipelines}}" and "\textbf{\textit{initialize\_}}\textbf{\textit{with\_readme}}".\end{tabular}  
&                      &                      & \includegraphics[scale=0.01]{figure/mark1.pdf}                  \\ 
 \cline{2-8}
& 6 & 1.~GET /api/v4/groups                                                                                                                                                                              & 1.~\texttt{500} Internal Error                                                                                     &  \begin{tabular}[c]{@{}c@{}}List groups with 2 parameters "\textbf{\textit{statistics}}" \\and~"\textbf{\textit{min\_access\_level}}".\end{tabular}                                                                                                                                                               & \includegraphics[scale=0.01]{figure/mark1.pdf}                 & \includegraphics[scale=0.01]{figure/mark1.pdf}                 & \includegraphics[scale=0.01]{figure/mark1.pdf}                  \\ 
 \cline{2-8} 
& 7 & \begin{tabular}[c]{@{}l@{}}1. GET /api/v4/groups\end{tabular}                                                                                                              & \begin{tabular}[c]{@{}l@{}}1.~\texttt{500} Internal Error\end{tabular}                             & Use <"\emph{\textbf{per\_page}}", "\emph{\textbf{0}}"> pair in the request.                                                                                                                                                                                                                   & \includegraphics[scale=0.01]{figure/mark1.pdf}                 & \includegraphics[scale=0.01]{figure/mark1.pdf}                 & \includegraphics[scale=0.01]{figure/mark1.pdf}                  \\ 
 \cline{2-8} 
& 8 & 1.~POST /api/v4/groups                                                                                                                                                                             & 1.~\texttt{500} Internal Error                                                                                     & \begin{tabular}[c]{@{}c@{}}Error when the parameter~"\textbf{\textit{praent\_id}}" is set\\ to be a special number like 2, -1, -2.~\end{tabular}                                                                                                                             & \includegraphics[scale=0.01]{figure/mark1.pdf}                 & \includegraphics[scale=0.01]{figure/mark1.pdf}                 & \includegraphics[scale=0.01]{figure/mark1.pdf}                  \\ 
 \cline{2-8} 
 & 9 &1.~POST /api/v4/groups                                                                                                                                                                             & 1.~\texttt{500} Internal Error                                                                                     & \begin{tabular}[c]{@{}c@{}}Create group with the undefined parameter\\ "\textbf{\textit{shared\_runners\_}}\textbf{\textit{setting}}".\end{tabular}  
 &                      &                      & \includegraphics[scale=0.01]{figure/mark1.pdf}                  \\ 
 \cline{2-8} 
 & 10  &  \begin{tabular}[c]{@{}l@{}}1.~POST /api/v4/groups\\2.~DELETE /api/v4/groups/:id\\3.~GET /api/v4/groups/:id/descendant\_groups\end{tabular}          
 &\begin{tabular}[c]{@{}l@{}}1. \texttt{201}~Created\\2.~\texttt{202} Accepted\\3.~\texttt{200} OK\end{tabular}        
 &\begin{tabular}[c]{@{}c@{}} Use after free. The third request successfully\\ accesses the descendant groups of a\\ deleted group. Private data leakage.   \end{tabular} 
 &                        &         \includegraphics[scale=0.01]{figure/mark1.pdf}               & \includegraphics[scale=0.01]{figure/mark1.pdf}                  \\ 
 \cline{2-8} 
 & 11 & \begin{tabular}[c]{@{}l@{}}1.~POST /projects/:id/issues\\2.~PUT /projects/:id/issues/:issue\_iid\end{tabular}                                                                                                                                                                               & \begin{tabular}[c]{@{}l@{}}1. \texttt{201}~Created\\2.~\texttt{500} Internal Error\end{tabular}                                                                                    & \begin{tabular}[c]{@{}c@{}}Update issue with an extra-long label.\end{tabular}                                                                                                    &     \includegraphics[scale=0.01]{figure/mark1.pdf}      &         \includegraphics[scale=0.01]{figure/mark1.pdf}               & \includegraphics[scale=0.01]{figure/mark1.pdf}                  \\ 
 \cline{2-8} 
 & 12  &  \begin{tabular}[c]{@{}l@{}}1.~POST /api/v4/groups\\2.~DELETE /api/v4/groups/:id\\3.~GET /api/v4/groups/:id\end{tabular}          
 &\begin{tabular}[c]{@{}l@{}}1. \texttt{201}~Created\\2.~\texttt{202} Accepted\\3.~\texttt{200} OK\end{tabular}        
 &\begin{tabular}[c]{@{}c@{}} Use after free. The third request lists a group \\ with the parameters "\textbf{\textit{with\_custom\_attributes}}"  \\ and "\textbf{\textit{with\_projects}}". Private data leakage.   \end{tabular}                                     &                        &         \includegraphics[scale=0.01]{figure/mark1.pdf}               & \includegraphics[scale=0.01]{figure/mark1.pdf}                  \\ 
 \cline{2-8} 
 & 13  & \begin{tabular}[c]{@{}l@{}}1.~POST /projects/:id/repository/branches\\2.~DELETE /projects/:id/repository/branches/:Br\\3.~GET /projects/:id/repository/branches/:Br\\4.~POST /projects/:id/repository/branches
 \end{tabular}                                                                                                                                                                               & \begin{tabular}[c]{@{}l@{}}1. \texttt{201}~Created\\2. \texttt{204}~No Content\\3. \texttt{404}~Not Found\\4.~\texttt{201} Created\end{tabular}                                                                                    & \begin{tabular}[c]{@{}c@{}}Use after free. The fourth request\\ successfully creates a branch while its \\ "\textbf{\textit{ref}}"  refers to the name of the deleted \\ branch. Private data leakage.   \end{tabular}                                                                                                          &                      &         \includegraphics[scale=0.01]{figure/mark1.pdf}               & \includegraphics[scale=0.01]{figure/mark1.pdf}                  \\ 
 \cline{2-8} 
 & 14 &1.~POST /api/v4/projects/:proj\_id/repository/branches  & 1.~\texttt{500} Internal Error                                                                                     & \begin{tabular}[c]{@{}c@{}}Create a new branch with \\ the undefined parameter "\textbf{\textit{sort}}".\end{tabular}  
 &                      &                      & \includegraphics[scale=0.01]{figure/mark1.pdf}                  \\ 

 \hline
\hspace{-0.25cm} Bugzilla & 15 & \begin{tabular}[c]{@{}l@{}}1. POST /rest/bugs\end{tabular}                                                                                                              & \begin{tabular}[c]{@{}l@{}}1.~\texttt{500} Internal Error\end{tabular}                             & Use invalid datetime value in the request.
 &  \includegraphics[scale=0.01]{figure/mark1.pdf}  &  \includegraphics[scale=0.01]{figure/mark1.pdf} & \includegraphics[scale=0.01]{figure/mark1.pdf}  \\
 \hline
  \vspace{-0.3cm}\multirow{2}{*}{\begin{tabular}[c]{@{}l@{}}Word-\\Press\end{tabular}}
  & 16 &  \begin{tabular}[c]{@{}l@{}}1.~POST /categories $\Rightarrow$ Category id: (cid)\\2.~POST /categories \\3.~DELETE /categories({cid})  \\4.~POST /categories
  \end{tabular}                      
                                                                                                                                                           & \begin{tabular}[c]{@{}l@{}}1. \texttt{201}~Created\\2. \texttt{201}~Created\\3. \texttt{200}~OK\\4.~\texttt{500} Internal Error\end{tabular}                                                                                    & 
                                                                                                                                      \begin{tabular}[c]{@{}c@{}} Use after free.  The parent of both the second \\ and forth requests refers to the same id  of the \\ category (cid). However, the fourth request \\triggers an error after (cid) gets deleted.  \end{tabular}                                                                                                              &  & \includegraphics[scale=0.01]{figure/mark1.pdf}  & \includegraphics[scale=0.01]{figure/mark1.pdf} \\
 \cline{2-8} 
 & 17 & \begin{tabular}[c]{@{}l@{}}1. POST /wp/v2/posts \end{tabular}                                                                                                              & \begin{tabular}[c]{@{}l@{}}1.~Timeout \end{tabular}                             &
 Connections get closed after timeout. 
&  & \includegraphics[scale=0.01]{figure/mark1.pdf} & \includegraphics[scale=0.01]{figure/mark1.pdf} \\
 \hline
 \multicolumn{5}{c}{Total}                                                                                                                                                                                                                                                                                                                                                                                                                                                                                                                                                                   & 7     & 13     & \textbf{17}            \\
 \toprule
 \end{tabular}
 \\\footnotesize{$^{\rm a}$A use-after-free error on cloud services means that the request accesses a deleted resource, leading to bypassing resource quotas and corrupting service state.} 
 \vspace{-1.0em}
 \end{table*}

  $\bullet$
  The \emph{DataDriven Checker} of \sysname finds many unique errors caused by  misuse of undefined parameters. 
  For instance, \sysname finds 43 unique errors in total with the \emph{DataDriven Checker} on GitLab via the 5 REST APIs. 
  \sysname also finds 6 unique errors that misuse undefined parameters on Bugzilla Bugs API. 
  The results demonstrate that 1) incorrect parameter usage errors commonly exist in various cloud services,  
  and 2) \sysname can effectively discover this kind of unique errors with our data-driven designs.

\subsection{Real Error Analysis}

To further evaluate the performance of each fuzzer on real error discovery, 
we manually analyze all the errors on GitLab, Bugzilla and WordPress 
reported by each fuzzer. 
Specifically, 
we filter out the errors that cannot be reproduced due to the change of server states as discussed in Section~\ref{sec:hardreproduce}. 
Then, we deduplicate the errors that are caused by the same reason. 
The details of real errors on three cloud services found by each fuzzer are shown in Table~\ref{table:realbug}. 

  $\bullet$
  The exploration space of real errors triggered by RESTler is fully covered and exceeded  by \sysnamepart and \sysname. 
  After error filtering and deduplicating, 
  \sysnamepart finds all the errors discovered by RESTler, 
  and finds 6 more unique errors missed by RESTler, 
  from which we can have the following conclusions. 
  1) Since RESTler's possible request combinations are covered by \sysnamepart, 
  the real errors triggered by RESTler also can be found by \sysnamepart; 
  2) Thanks to the \emph{length-orientated sequence construction} implemented in \sysnamepart, 
  \sysnamepart assigns more execution times to long request sequences. 
  As a result, \sysnamepart finds 5 new unique errors triggered by 
  the request sequences of length greater than 2; 
 3) We find that RESTler cannot find the No. 16 error triggered by 4 requests, 
   but all four requests had been successfully generated by RESTler, 
as the number of unique request templates successfully generated by RESTler is the same as \sysnamepart and \sysname on WordPress Categories API as shown in Table~\ref{table:result}. 
   Thus, either RESTler fails to construct a sequence of sufficient length, 
or RESTler does not generate high-quality requests to pass the checking of WordPress. 
   This case study demonstrates the motivation of our paper, i.e., a new solution is demanded to improve the sequence construction and request generation.

$\bullet$
  The \emph{DataDriven Checker} implemented in \sysname can efficiently find the unique errors caused by incorrect parameter usage. 
  For instance, as shown by the No. 2 error, \sysname adds an undefined parameter "\textbf{\textit{initialize\_with\_readme}}" to 
   the \texttt{PUT} request, 
  which is the parameter used in other requests like \texttt{POST}. 
  Violating the standard API specification, GitLab executes the behavior of "\textbf{\textit{initialize\_with\_readme}}". 
  However, this behavior is not compatible with the behavior of \texttt{PUT}, 
  and causes the response of \texttt{500} Internal Error. 
  For another example shown by the No. 9 error, 
  \sysname sends a \texttt{POST} request with the undefined parameter "\textbf{\textit{shared\_runners\_setting}}". 
  When creating a group, GitLab accesses a non-existent resource because of this parameter, which causes the crash of the creation process. 
  As a result, GitLab returns the response of \texttt{500} Internal Error. 
  Although \sysname finds several incorrect parameter usage errors on Bugzilla and WordPress, we find that their root causes are the
  same as the shown errors in Table~\ref{table:realbug}. Thus, these errors are not shown in Table~\ref{table:realbug}. 

\begin{table*}\scriptsize
  \setlength{\abovecaptionskip}{0.1cm}
  \centering
  \caption{The pass rate and the number of unique errors found by RESTler, RESTler+Seq, RESTler+Rec1, RESTler+RecList, RESTler+Model, and \sysnamepart, respectively. } 
  \label{table:stepwise} 
  \begin{tabular}{cccc|cc|cc|cc|cc|cc} 
  \toprule
  \multirow{2}{*}{Cloud Service}    & \multirow{2}{*}{REST API~}  & \multicolumn{2}{c|}{RESTler} & \multicolumn{2}{c|}{RESTler+Seq}  & \multicolumn{2}{c|}{RESTler+Rec1}  & \multicolumn{2}{c|}{RESTler+RecList}  & \multicolumn{2}{c|}{RESTler+Model}  & \multicolumn{2}{c}{\sysnamepart} \\
  \cline{3-14}
                             &                        & Pass Rate       & Errors   & Pass Rate      & Errors   & Pass Rate    & Errors    & Pass Rate       & Errors     & Pass Rate       & Errors & Pass Rate       & Errors       \\ 
  \toprule
  \multirow{3}{*}{GitLab}    & Projects               & 69.48\%     & 7.00    & 89.57\% & 7.50 & 66.10\% & 7.50 & 65.66\% & 7.25 & 87.11\%    & 9.00 & 94.70\%  & 9.75\\ 
                             & Groups                 & 60.28\%     & 16.00   & 94.67\%  &  19.50 & 65.22\% & 19.00 & 65.56\% & 17.00 & 94.53\%    & 19.00& 94.66\%& 19.50        \\
                             & Issues                 & 86.32\%     & 6.50    & 89.06\% & 6.50  & 88.14\% & 7.00  & 90.38\% & 7.00 &91.00\% & 7.50  & 94.31\%& 7.00           \\ 
                             \hline
  \multirow{1}{*}{Bugzilla}  & Comments               & 86.19\% & 8.00 & 89.05\% & 8.00 & 87.05\% & 8.00 & 89.05\% & 8.00 & 87.94\% & 8.25 & 89.36\% & 8.00 \\ 
  \hline
  \multirow{1}{*}{WordPress} & Categories             & 81.35\% & 9.00  & 86.40\% & 8.75 & 81.03\%  & 9.00& 81.44\% & 9.00 & 82.41\% & 9.50 & 90.53\% & 9.50 \\
  \toprule
  \end{tabular}
  \vspace{-2.0em}
  \end{table*}

\noindent \textbf{Security bugs.} 
Our fuzzers find 5 security bugs 
that allow attackers to access resources that have been deleted, 
which may cause the leakage of private user data. 
For instance, as shown by the No. 13 error, 
the fourth request creates a new branch referring to the deleted one, 
which copies the resource data from the deleted one. 
Therefore, attackers can leverage the No. 13 error to restore the deleted private data of users. 
These cases demonstrate the significant performance of \sysname 
in constructing long request sequences to explore head-to-reach states and find security bugs.
On the contrary,  RESTler hardly generates a request sequence 
of length greater than 2, 
and cannot trigger any of the 5 security bugs.

  We have reported all the discovered unique errors on GitLab, 
  Bugzilla and WordPress, 
  and 7 of them have already been confirmed as logic bugs by vendors.

\section{Further Analysis}\label{sec:analysis}

\subsection{Stepwise Analysis}\label{sec:stepwise}
In order to evaluate the contribution of our data-driven designs 
to the pass rate and error discovery, 
we construct the following 4 fuzzers with different designs. 
To be specific, 
we construct {RESTler+Seq} by implementing the \emph{length-orientated sequence construction} in RESTler;
{RESTler+Rec1} is the one that generates requests by randomly 
using a \textit{param-value pair} recorded in the \texttt{Collection Module} 
and using default values for other parameters; 
{RESTler+RecList} randomly uses a list of recorded \textit{param-value pairs} to generate requests, 
i.e., it reproduces and sends the same requests to trigger the same behaviors of cloud services;
and we combine RESTler with the \emph{attention model-based request generation} to construct {RESTler+Model}. 
Then, we can measure the contribution of the \emph{length-orientated sequence construction} 
by comparing the results of \sysnamepart and RESTler+Model, and the results of RESTler+Seq and RESTler.   
We can measure the contribution of the \emph{attention model-based request generation} 
from the results of \sysnamepart and RESTler+Seq, and the results of RESTler+Model and RESTler. 
Furthermore, we can evaluate the contribution of different request generation algorithms 
on pass rate and error discovery 
by comparing the performance of RESTler+Rec1, RESTler+RecList and RESTler+Model. 
Since the evaluation in Section~\ref{sec:eval} demonstrates 
the significant performance of the \emph{DataDriven Checker} 
on the discovery of incorrect parameter usage error, 
we do not evaluate its performance in this subsection. 
Each evaluation lasts for 12 hours and is repeated 4 times. 
The experiment settings are the same as in Section~\ref{section:settings}, 
and the results are shown in Table~\ref{table:stepwise}, 
from which we have the following conclusions.

$\bullet$
The \emph{length-orientated sequence construction} significantly improves the pass rate of a REST API fuzzer, 
while it cannot improve the error discovery on most evaluations. 
For instance,  the pass rate of \sysnamepart 
is 9.85\% higher than  RESTler+Model on WordPress Categories API,  
and the average pass rate of RESTler+Seq 
is 26.48\% higher than RESTler on GitLab. 
On the other hand, RESTler+Seq performs closely to RESTler on error discovery. 
To figure out the reason, 
we analyze the sequence templates constructed by RESTler+Seq, 
and have the following conclusion. 
The sequence templates of RESTler+Seq are mainly composed of easy-to-construct request templates, 
e.g., \texttt{GET}, 
which contain few parameters and are easy to pass the cloud service's checking. 
Because of the \emph{length-orientated sequence construction}, 
RESTler+Seq tends to generate sequences based on long sequence templates, 
in which all the requests have passed the checking. 
However, due to low request generation quality, 
RESTler+Seq rarely stores hard-to-construct request templates within its sequence templates, 
which often consist of easy-to-construct request templates. 
As a result, the executions of easy-to-construct requests increase by dozens of times, which results in a high pass rate. 
On the contrary, \sysnamepart improves the pass rate of hard-to-construct requests 
with the help of the \emph{attention model-based request generation}, 
which improves the diversity of sequence templates and error discovery.


\begin{table*}[t]\scriptsize
    \setlength{\abovecaptionskip}{0.1cm}
  \centering
  \caption{The published bugs found by RESTler and \sysname on GitLab.}
  \label{table:realbug_appendix}
  \begin{tabular}{m{0.8cm}<{\centering} llc m{2.1cm}<{\centering} m{1.5cm}<{\centering} } 
  \toprule
  No.~ & \multicolumn{1}{c}{Endpoints of Used Requests}                                                                                                                                                     & \multicolumn{1}{c}{Responses}   &  Description    & RESTler       & \sysname    \\ 
  \toprule
  1    & \begin{tabular}[c]{@{}l@{}}1. POST /api/v4/projects\\2. GET /api/v4/projects/:id/custom\_attributes\\3. DELETE /api/v4/projects/:id\\4. GET /api/v4/projects/:id/custom\_attributes\\\end{tabular} & \begin{tabular}[c]{@{}l@{}}1.~\texttt{201} Created\\2. \texttt{200} OK\\3.~\texttt{202} Accepted\\4.~\texttt{500} Internal Error\end{tabular}   & \begin{tabular}[c]{@{}c@{}} Use after free $^{\rm a}$. The fourth request \\tries to access the attributes of a \\deleted project. \end{tabular}  & & \includegraphics[scale=0.008]{figure/mark1.pdf}   \\ 
  \hline
  2    & \begin{tabular}[c]{@{}l@{}}1. POST /api/v4/projects\\2. POST /api/v4/projects/:id\_1/fork\\3. POST /api/v4/projects\\4. POST /api/v4/projects/:id\_2/fork\\5. POST /projects/:id\_1/fork/:id\_2\\ \end{tabular} & \begin{tabular}[c]{@{}l@{}}1.~\texttt{201} Created\\2.~\texttt{201} Created\\3.~\texttt{201} Created\\4.~\texttt{201} Created\\5.~\texttt{500} Internal Error\end{tabular} & \begin{tabular}[c]{@{}c@{}}Error when creating a forked\\ relationship between existing projects\\ that already have forks.\end{tabular} && \includegraphics[scale=0.008]{figure/mark1.pdf}     \\ 
  \hline
  3    & 1.~GET /api/v4/groups                                                                                                                                                                              & 1.~\texttt{500} Internal Error &\begin{tabular}[c]{@{}c@{}}List groups with 2 parameters "\textbf{\textit{statistics}}" \\and~"\textbf{\textit{min\_access\_level}}".\end{tabular}& \includegraphics[scale=0.008]{figure/mark1.pdf}    & \includegraphics[scale=0.008]{figure/mark1.pdf}     \\ 
  \hline
  4    & 1. GET /api/v4/groups            & 1.~\texttt{500} Internal Error     &Use <"\emph{\textbf{per\_page}}", "\emph{\textbf{0}}"> pair in request.       & \includegraphics[scale=0.008]{figure/mark1.pdf}    & \includegraphics[scale=0.008]{figure/mark1.pdf}         \\ 
  \hline
  \multicolumn{4}{c}{Total}                                                                                                                                                                                                                                                                                                                                                                                                                                                                                                                                                                   & 2       & \textbf{4}            \\
  \toprule
  \end{tabular}
  \\\footnotesize{$^{\rm a}$A use-after-free error on a cloud service means that the request accesses a deleted resource.}
      \vspace{-2.0em}
  \end{table*}
  
  \begin{table}[t]\scriptsize
    \setlength{\abovecaptionskip}{0.1cm}
  \centering
  \caption{The average time spent by RESTler and \sysname to trigger the published bugs on GitLab, where N/A means that the fuzzer fails to trigger the bug in 48 hours.}\label{table:reproduce}
  \begin{tabular}{m{0.2cm}<{\centering} m{0.5cm}<{\centering} m{1.2cm}<{\centering} m{1.8cm}<{\centering} m{1.0cm}<{\centering} m{1.2cm}<{\centering}} 
  \toprule
  No. & REST API      &   \# Request Templates & Minimum  Length  to Trigger Bugs & RESTler  & \sysname \\ \hline
  1       & Projects & 3   & 3    & N/A     & 66.5 mins   \\ \hline
  2       & Projects & 4   & 5    & N/A     & 142.6 mins  \\ \hline
  3       & Groups   & 1   & 1    & 27.6 mins     & 26.6 mins   \\ \hline
  4       & Groups   & 1   & 1    & 13.7 mins     & 14.2 mins   \\ \toprule
  \end{tabular}
  \vspace{-2.3em}
  \end{table}

$\bullet$
The \emph{attention model-based request generation} can improve the error discovery and pass rate of a REST API fuzzer. 
For instance, RESTler+Model finds more errors than RESTler on all 5 evaluations. 
\sysnamepart also performs better than RESTler+Seq on pass rate and error discovery  
with the help of the \emph{attention model-based request generation}. 
The results demonstrate the contribution of our new design to the fuzzing performance. 
Furthermore, \sysnamepart performs the best on pass rate and error discovery on most evaluations, 
which implies that the two data-driven designs can be combined to improve each other.

$\bullet$
The \emph{attention model-based request generation} can effectively provide  key \textit{param-value lists} for request generation 
and improve the fuzzing performance compared to straightforward algorithms. 
For instance, the pass rate and error discovery of RESTler+Model is better than RESTler+Rec1 
on most evaluations. 
The error discovery of RESTler+Model is significantly better than RESTler+RecList on all 5 evaluations. 
Thus, neither randomly selecting a mutation nor simply reproducing a well-constructed request 
can  improve the performance of a REST API fuzzer. 
The results demonstrate that our design learns the implicit relations between key \textit{param-value pairs}, 
and generates diverse yet valid mutations as \textit{param-value lists}. 
Our design can serve as a new direction 
to improve request generation quality for REST API fuzzing.

\subsection{Performance on Reproducing Serious Bugs}

To evaluate the performance of RESTler and \sysname on serious bug discovery, 
we manually check the published issues of GitLab, 
and collect 4 serious bugs that a REST API fuzzer can reproduce as the ground truth. 
Then, we construct the fuzzing grammar that only contains the relevant request templates 
for each unique bug, 
as shown in Table~\ref{table:reproduce}. 
For instance, to reproduce the  No. 2 bug, 
a fuzzer needs to correctly generate  5 requests in a reasonable order, 
which are based on 4 different request templates in the grammar. 
Each evaluation lasts up to 48 hours and is repeated 5 times to calculate the average trigger time. 
The results are shown in Table~\ref{table:realbug_appendix} and Table~\ref{table:reproduce}, 
from which we have the following observations.

$\bullet$
\sysname can reproduce 
the serious bugs published on GitLab more effectively and efficiently compared to RESTler. 
\sysname can reproduce all  4  bugs, while RESTler can reproduce two. 
Furthermore, it takes 142.6 mins on average for \sysname 
to construct a sequence with 5 requests in the correct order 
as shown in Table~\ref{table:realbug_appendix}, 
 while RESTler  cannot construct the  sequence within 48 hours. 
The results demonstrate the significant performance of \sysname on serious bug discovery, 
and we analyze the reasons as follows. 
First, \sysname can construct long request sequences with the \emph{length-orientated sequence construction}, which increases the execution times on complicated request combinations compared to RESTler. 
Second, the \emph{attention model-based request generation} implemented in \sysname improves the pass rate of hard-to-construct requests to GitLab, 
which reduces the trigger time of bugs. 

$\bullet$
Our designs will not prevent \sysname from triggering relatively simple bugs. 
As shown in Table~\ref{table:realbug_appendix},  
 the No. 3 and No. 4 bugs are triggered by one request with incorrect parameters and values. 
 Both \sysname and RESTler trigger the two bugs by 
 using the traditional generation method to generate the corresponding request,  
 and the average time spent by the two fuzzers is close. 
 In conclusion, \sysname performs closely with RESTler on discovering this kind of bugs, 
 which are triggered by one incorrectly constructed request.   

\begin{figure*}[t]
  \setlength{\abovecaptionskip}{0.3cm}
  \setlength{\belowcaptionskip}{-0.6cm}
      \centering
      \begin{subfigure}[b]{0.33\textwidth}    
          \centering
          \includegraphics[width=2.3in,height=1.7in]{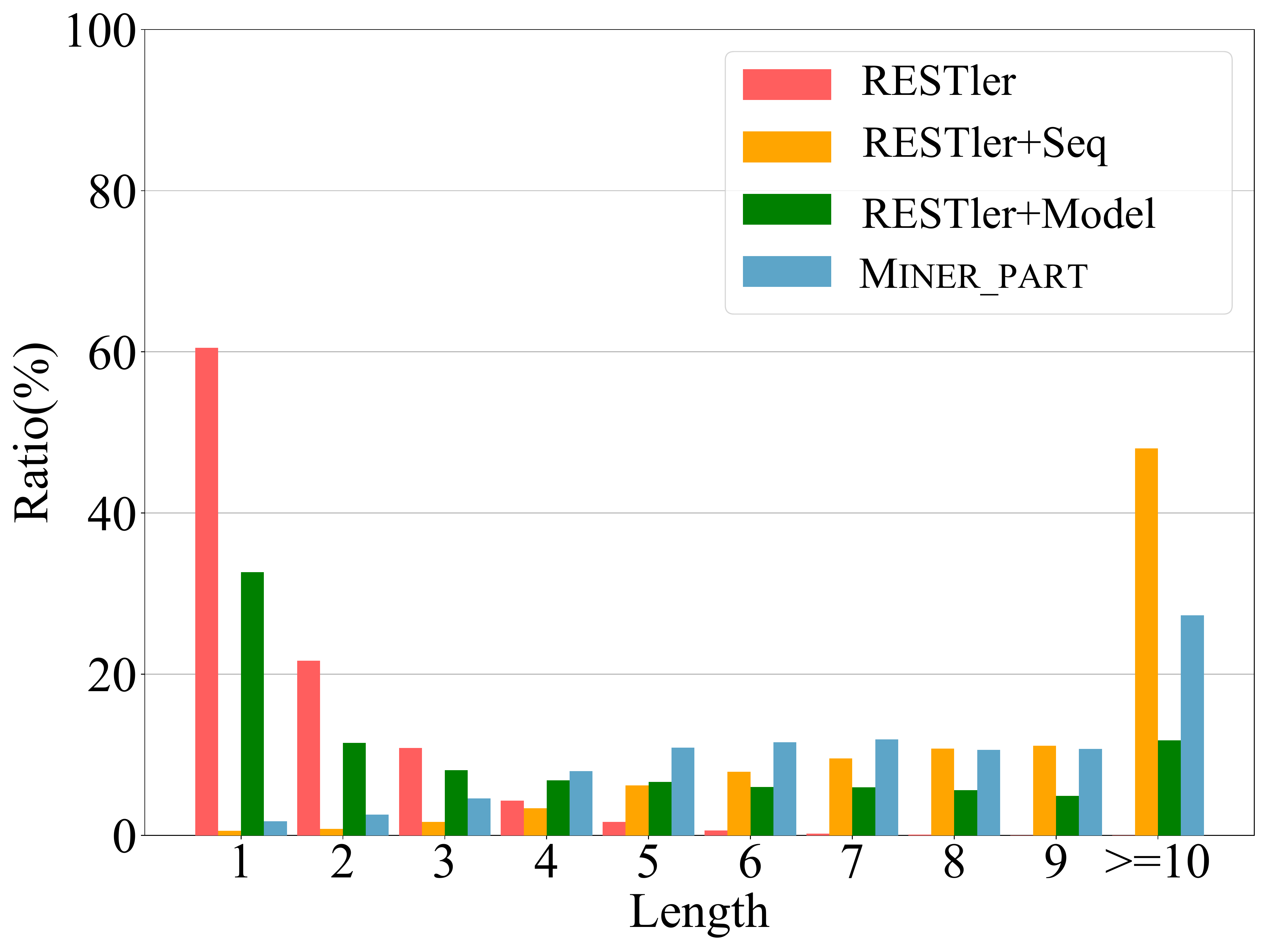}
           \setlength{\abovecaptionskip}{-0.45cm}
          \caption{GitLab Projects API.}
      \end{subfigure}
      \centering
         \begin{subfigure}[b]{0.33\textwidth}
             \centering
             \includegraphics[width=2.3in,height=1.7in]{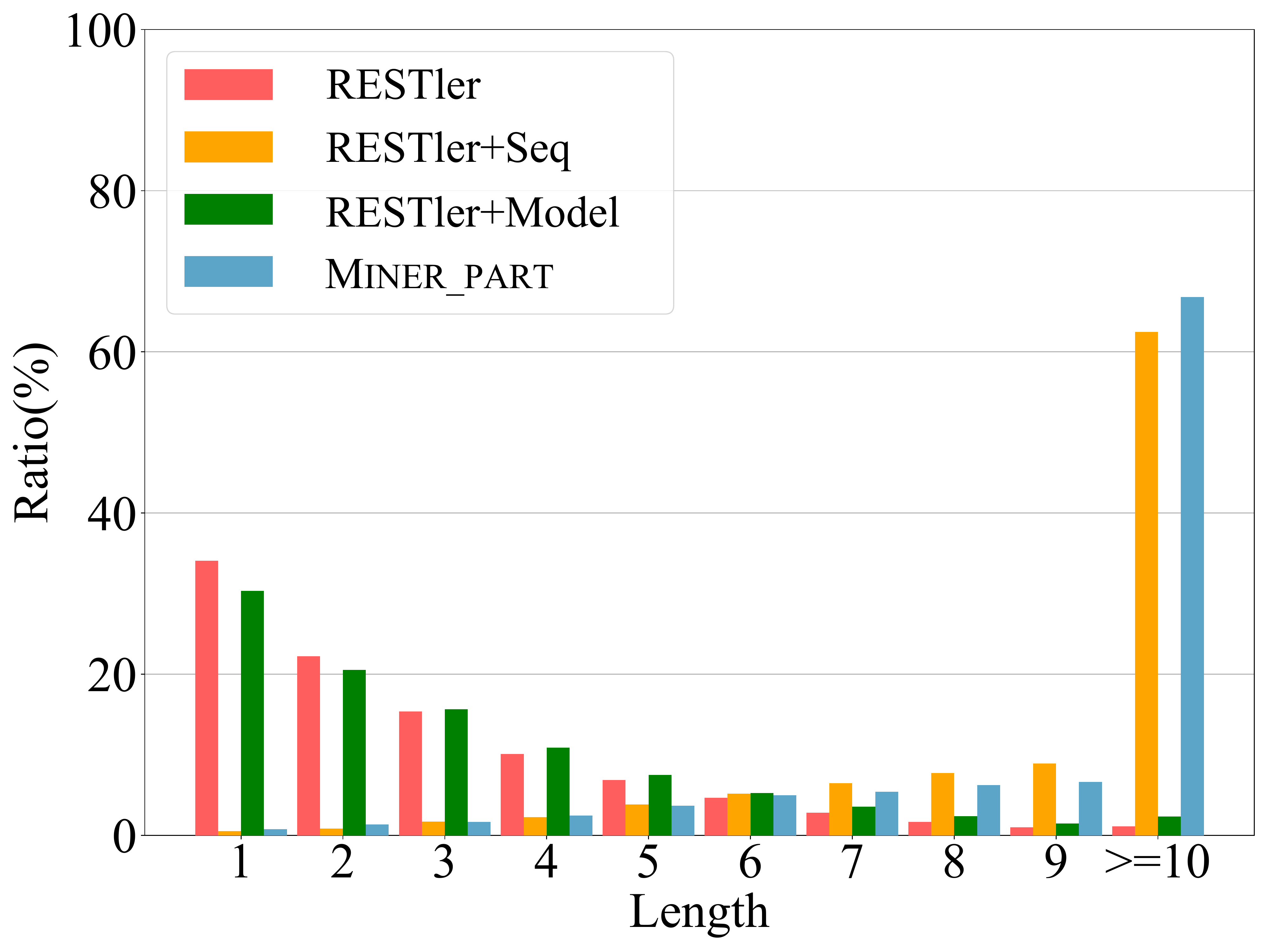}
             \setlength{\abovecaptionskip}{-0.45cm}
             \caption{Bugzilla Comments API.}
         \end{subfigure}
         \begin{subfigure}[b]{0.33\textwidth}
             \centering
             \includegraphics[width=2.3in,height=1.7in]{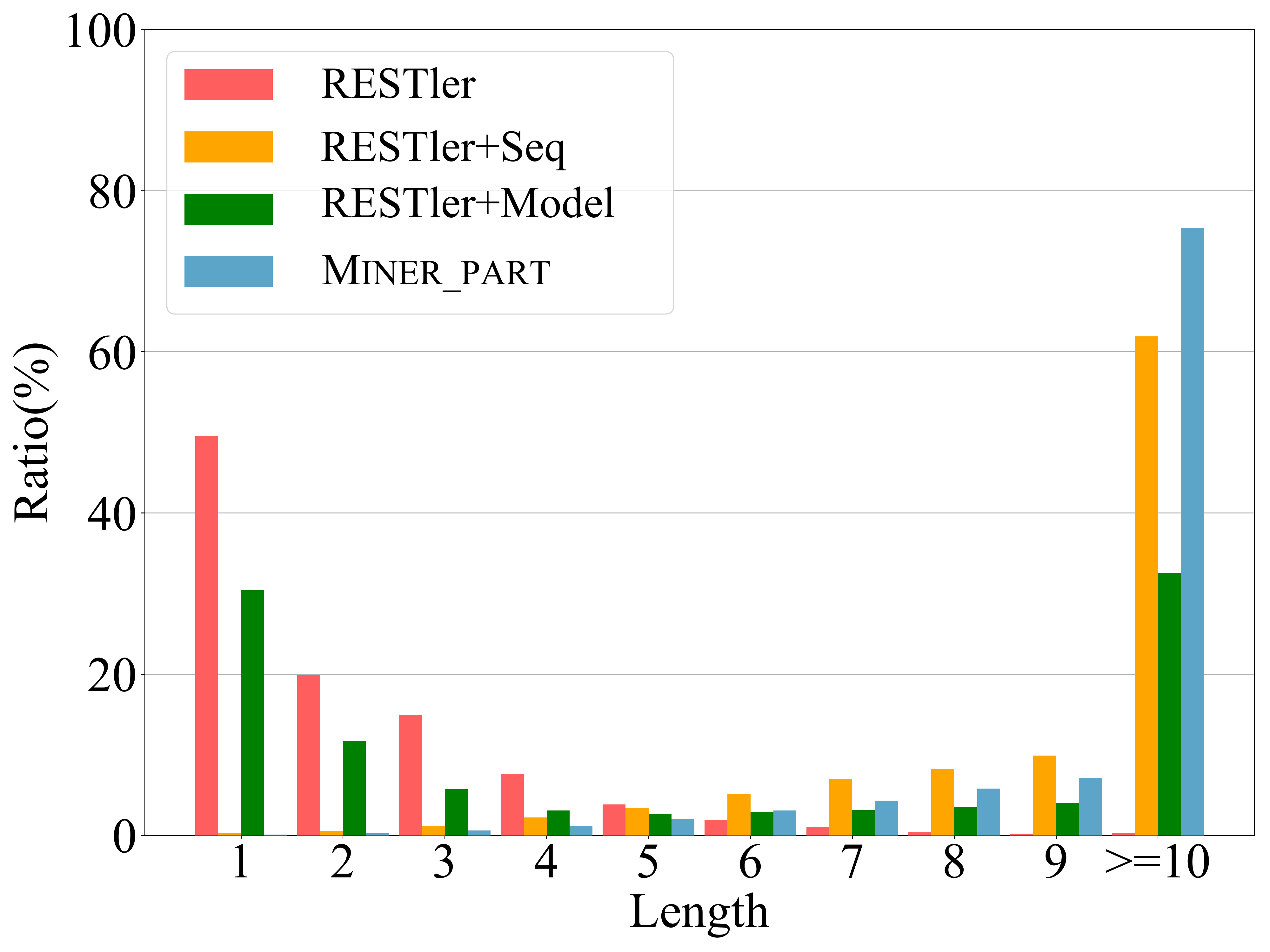}
             \setlength{\abovecaptionskip}{-0.45cm}
             \caption{Wordpress Categories API.}
         \end{subfigure}
         \vspace{-0.9mm}
 
             \caption{The distribution of  sequence length sent by 4 fuzzers, where RESTler+Seq only implements the \emph{length-orientated sequence construction} and RESTler+Model only implements the \emph{attention model-based request generation}.} 
             \label{fig:seqlength} 
      
 \end{figure*}

\vspace{-4mm}
\subsection{Sequence Length Analysis}\label{sec:sequencelengthanalysis}

To analyze the contribution of different designs to sequence length, 
we conduct the experiments as follows. 
We evaluate RESTler, RESTler+Seq, RESTler+Model, and \sysnamepart on GitLab Projects API, 
Bugzilla Comments API, and WordPress Categories API. 
Each evaluation lasts for 12 hours, during which we count the executions of sequences with different lengths. 
Based on the above settings, 
we can measure the contribution of the \emph{attention model-based request generation} to sequence length from the results of RESTler and RESTler+Model, and the results of RESTler+Seq and \sysnamepart. 
We can measure the contribution of the \emph{length-orientated sequence construction} to sequence length from the results of RESTler and RESTler+Seq, and the results of RESTler+Model and \sysnamepart. 
Since the \emph{DataDriven Checker} does not affect the sequence extension, 
we omit \sysname in the evaluation. 
The results are shown in Fig.~\ref{fig:seqlength}, from which we have the following conclusions.

$\bullet$
As shown in Fig.~\ref{fig:seqlength}, the \emph{attention model-based request generation} can increase the executions of long request sequences. For instance, RESTler+Model assigns more executions on longer request sequences than RESTler on all 3 targets. \sysnamepart also generates more request sequences with larger lengths compared to RESTler+Seq. The results demonstrate the contribution of the \emph{attention model-based request generation} to sequence extension. By improving the pass rate of requests with our design, a REST API fuzzer can construct request sequences with larger lengths.

$\bullet$
The \emph{length-orientated sequence construction} significantly increases the executions of request sequences with large lengths. 
Both \sysnamepart and RESTler+Seq assign most executions to the request sequences whose lengths are greater than 10, demonstrating our design's effectiveness. 

\vspace{-2mm}
\subsection{Coverage Performance Analysis}\label{sec:coverage}

\begin{figure*}[t]
    \setlength{\abovecaptionskip}{0.2cm}
    \setlength{\belowcaptionskip}{-0.3cm}
     \centering
     \begin{subfigure}[b]{0.32\textwidth}    
         \centering
         \includegraphics[width=2.35in,height=1.8in]{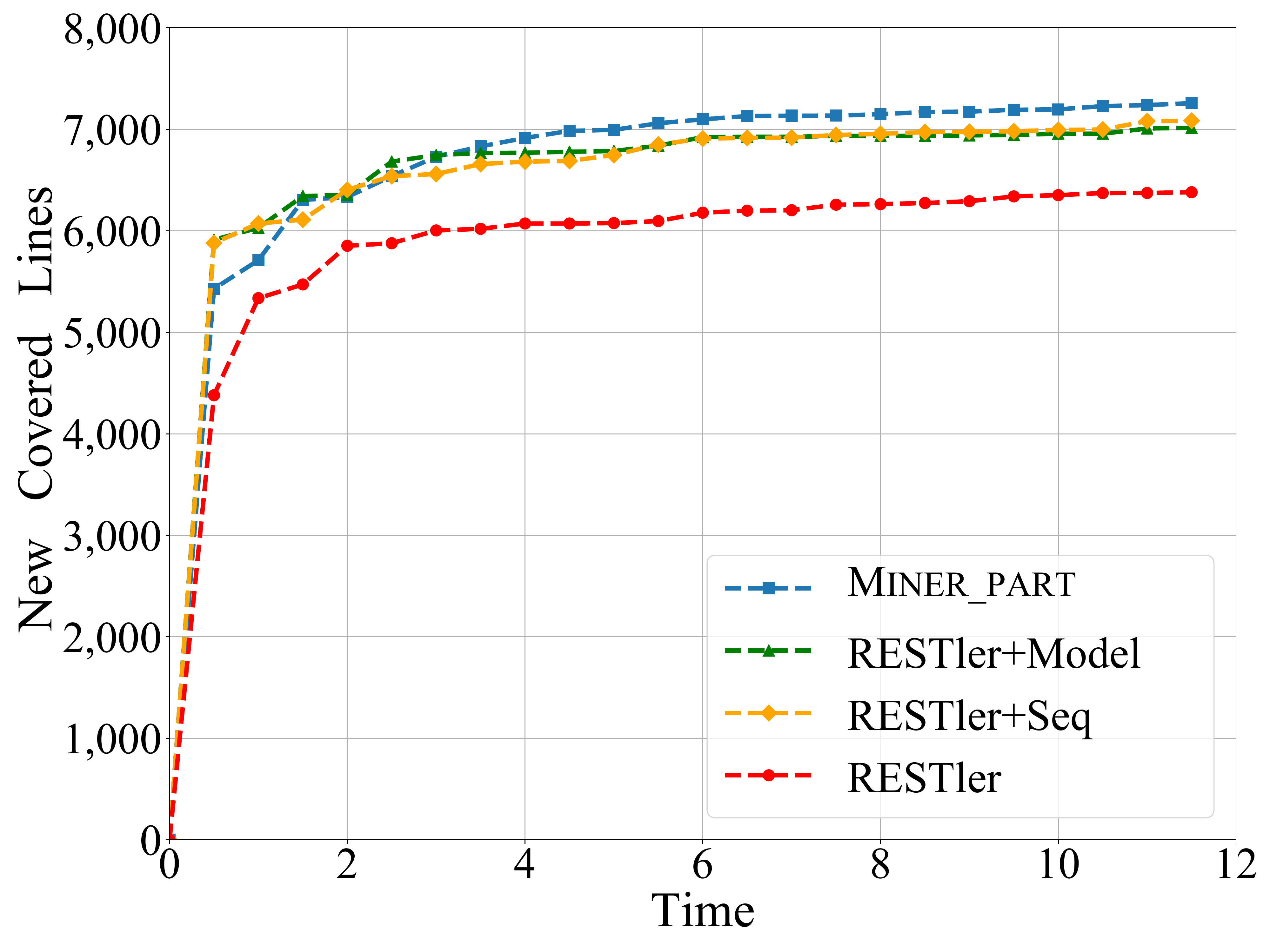}
         \setlength{\abovecaptionskip}{-0.4cm}
         \caption{GitLab Projects API.}
     \end{subfigure}
     \hfill
     \begin{subfigure}[b]{0.32\textwidth}
         \centering
         \includegraphics[width=2.35in,height=1.8in]{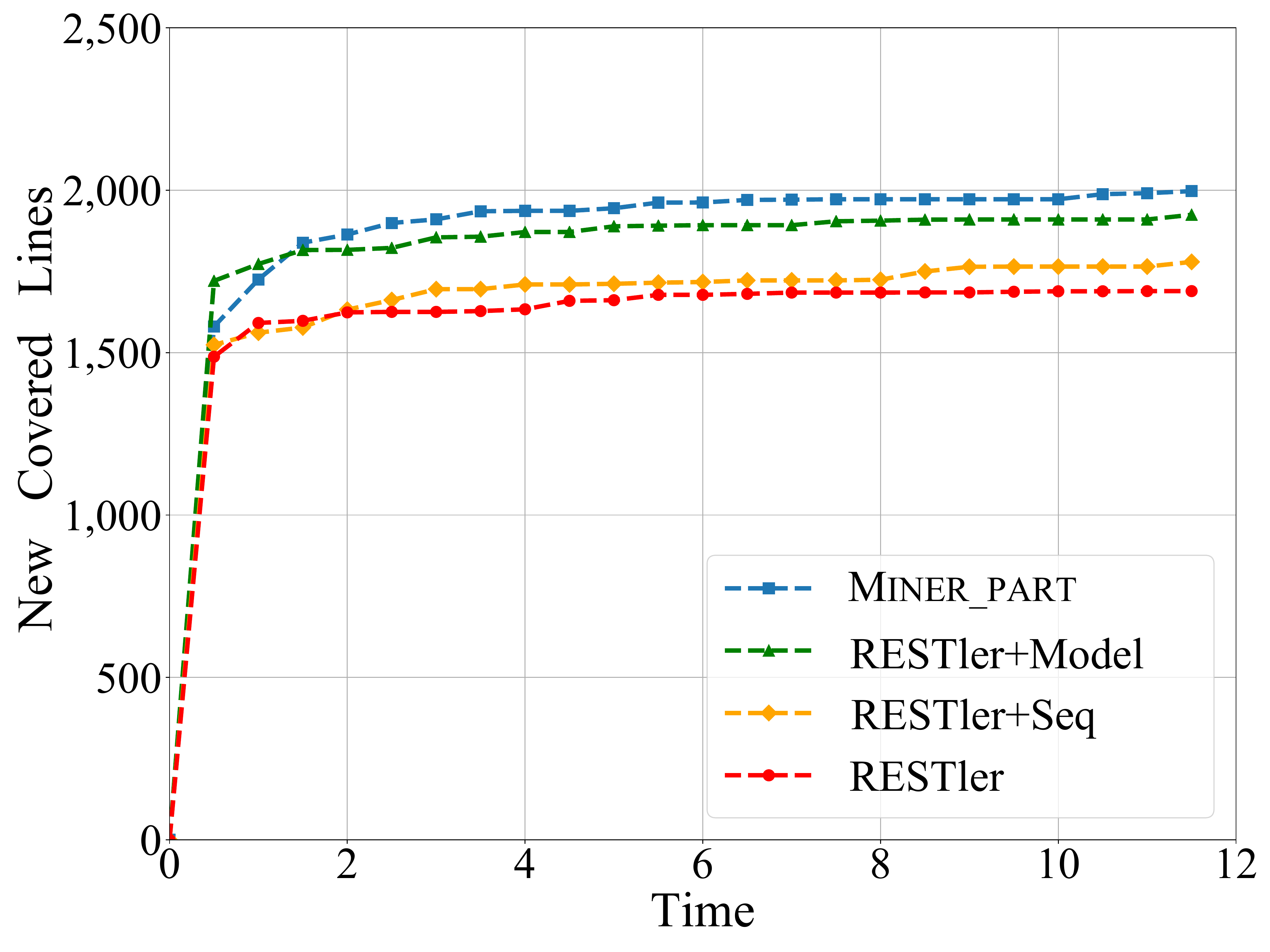}
         \setlength{\abovecaptionskip}{-0.4cm}
         \caption{GitLab Groups API.}
     \end{subfigure}
     \hfill
     \begin{subfigure}[b]{0.32\textwidth}
         \centering
         \includegraphics[width=2.35in,height=1.8in]{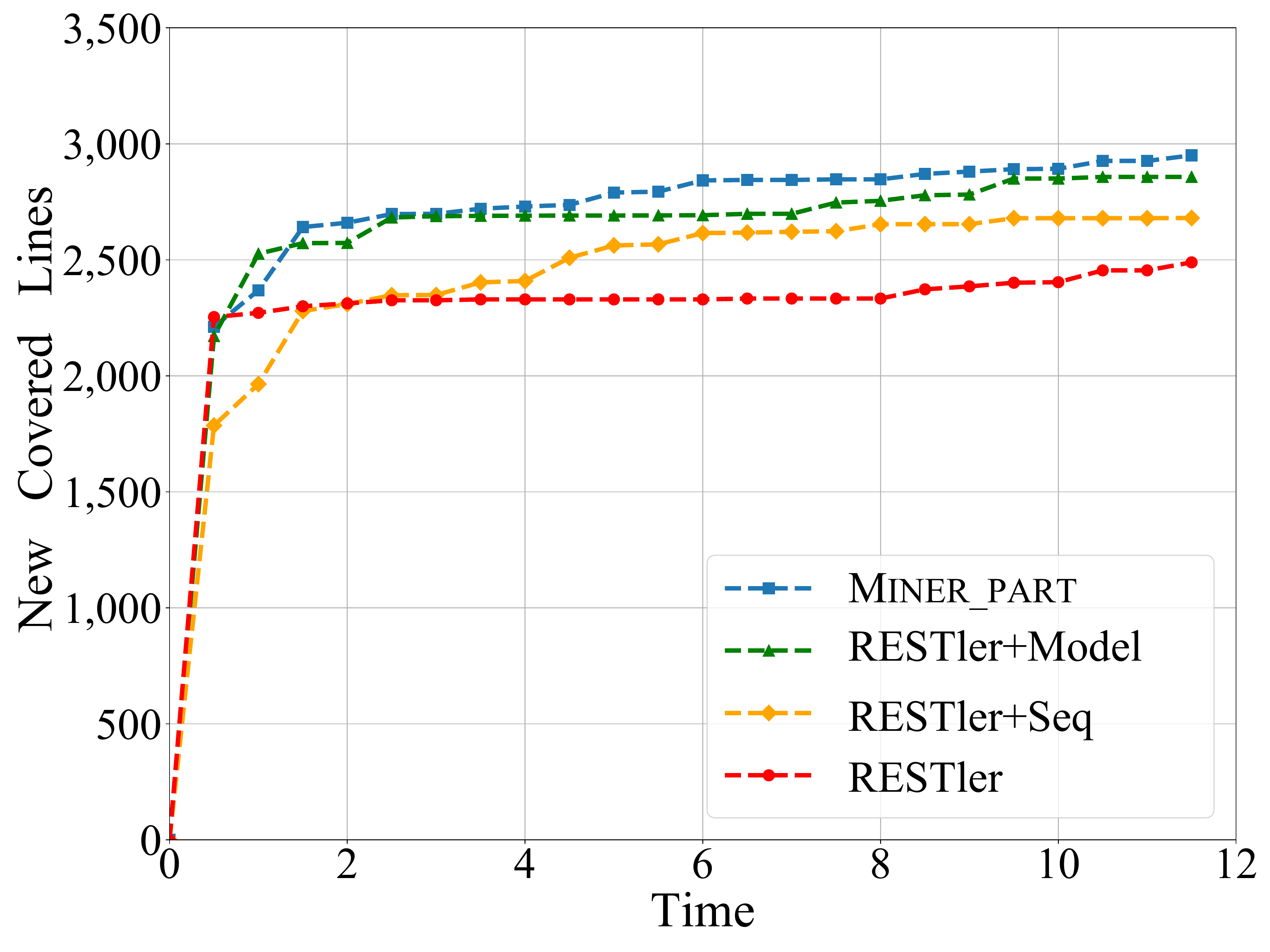}
         \setlength{\abovecaptionskip}{-0.4cm}
         \caption{GitLab Issues API.}
     \end{subfigure}
     \hfill
     \vspace{0.1cm}
            \caption{The line coverage growth of  4 fuzzers when fuzzing GitLab through  3 REST APIs, respectively.}
            \label{fig:coveredlines}
\end{figure*}

To evaluate the coverage performance of each fuzzer, 
we hook the source code of GitLab to trace the number of covered code lines triggered by the incoming requests. 
Specifically, we leverage a popular ruby gem named \emph{Coverband}~\cite{coverband} to trace the code lines executed in GitLab's \texttt{service/lib} and \texttt{service/app} folders. 
We also exclude the coverage statistics for irrelevant behaviors like service boot. 
Then, we conduct the line coverage evaluation for RESTler, RESTler+Seq, RESTler+Model, and \sysnamepart, 
each of which lasts for 12 hours. 
Since the \emph{DataDriven Checker} reuses the \textit{param-value pairs}
collected from the used requests and triggers the repeated code lines, 
it does not affect line coverage. 
Thus, we omit \sysname in the evaluation. 
The line coverage growth of each fuzzer is shown in Fig.~\ref{fig:coveredlines}, 
from which we have the following observations.

$\bullet$
Both the \emph{attention model-based request generation} and \emph{length-orientated sequence construction} 
can improve line coverage. For instance, the line coverage of RESTler+Model and
RESTler+Seq is better than RESTler on all the evaluations. 
Furthermore, the line coverage of RESTler+Model is better than RESTler+Seq on most targets, which implies that a sequence with well-constructed requests triggers deeper states of a cloud service than a long sequence with easy-to-construct requests. 

\begin{table*}[t]\scriptsize
  \setlength{\abovecaptionskip}{0.1cm}
  \setlength{\belowcaptionskip}{-0.1cm}
\centering
\caption{The average time overhead of the attention model and 
the pass rate of requests sent by \sysnamepart when the iteration duration of the \texttt{Training Module} is 1 hour, 2 hours and 3 hours, respectively.
} \label{table:passrate}
\begin{tabular}{C{3cm} C{2cm} C{1.5cm}|C{2cm}  C{1.5cm}|C{2cm}  C{1.5cm}} 
\toprule
\multirow{2}{*}{Target} & \multicolumn{2}{c|}{1 Hour} & \multicolumn{2}{c|}{2 Hours} & \multicolumn{2}{c}{3 Hours}  \\ 
\cline{2-7}
               & Time Overhead  & Pass Rate  & Time Overhead    & Pass Rate  & Time Overhead     & Pass Rate     \\ 
\toprule
GitLab Projects API    & {346.20s}  &   {92.78\%}        &  {380.80s}        &  {91.54\%}            &  {989.30s}         &    {89.55\%}            \\ 
\hline
GitLab Groups API     &  318.90s     & 93.49\%            & 386.90s & 92.80\%                &   1,002.67s   & 91.92\%                  \\ 
\hline
GitLab Issues API      & {463.40s}      & {90.90\%}            & {498.67s}          & {90.67\%}               & {1,001.60s}       &{90.23\%}   \\
\hline
Bugzilla Comments API & 97.60s & 88.89\% & 117.30s & 86.47\%	& 143.70s	& 85.37\% \\
\hline
WordPress Categories API & 305.93s &	91.02\%	 & 323.25s &	90.84\%	 & 402.40s & 90.14\% \\
\bottomrule
\end{tabular}
\vspace{-2.0em}
\end{table*}

$\bullet$
\sysnamepart performs the best on line coverage on all the evaluations. For instance, \sysnamepart covers the most code lines at the end of each evaluation.  The line coverage triggered by \sysnamepart grows rapidly on GitLab Groups API and GitLab Issues 
API. The results demonstrate that our data-driven designs can be combined to improve the coverage performance of a REST API fuzzer.



\vspace{-1mm}
\subsection{Schedule of the \texttt{Training Module}}\label{sec:overheadandpassrate}

To evaluate the impact of different schedules of the \texttt{Training Module} on fuzzing performance, we measure the average time overhead and pass rate for \sysnamepart when using 1 hour, 2 hours and 3 hours as the iteration duration of the \texttt{Training Module}, respectively. Each evaluation lasts for 12 hours. Thus, the invocation times of the \texttt{Training Module} are 12, 6 and 4 when using the above iteration durations, respectively. For each evaluation, we 1) calculate the average time overhead for each invocation of the \texttt{Training Module} to train the attention model and generate the \textit{param-value lists}, and 2) calculate the pass rate of requests. The results are shown in Table~\ref{table:passrate}, from which we have the following conclusions. Note that due to the different number of request templates and the different number of parameters in each request template, the number of the collected \textit{param-value pairs} (i.e., the training data for the attention model) is different when using \sysnamepart to fuzz a cloud service via different REST APIs. Thus, the average time overhead to train the attention model and generate \textit{param-value lists} can be different when fuzzing different targets.

$\bullet$
The average time overhead to train the attention model and generate \emph{param-value lists} increases when using a larger iteration duration of the \texttt{Training Module}. For instance, as shown in Table~\ref{table:passrate}, the average time overhead of the \texttt{Training Module} increases by 185.76\% when using 3 hours as the iteration duration to fuzz GitLab Projects API compared to 1 hour. The average time overhead  increases by 100.85\% when using 3 hours as the iteration duration to fuzz GitLab Issues API compared to 2 hours. We analyze the reasons for the results as follows. When the iteration duration of the \texttt{Training Module} increases, the training data collected in the \texttt{Collection Module}  increases correspondingly. Thus, the training set of the attention model increases, which requires more time to finish the training. 

\begin{figure*}
    \setlength{\abovecaptionskip}{0.2cm}
    \setlength{\belowcaptionskip}{-0.5cm}
        \centering
        \begin{subfigure}[b]{0.33\textwidth}    
            \centering
            \includegraphics[width=2.3in,height=1.45in]{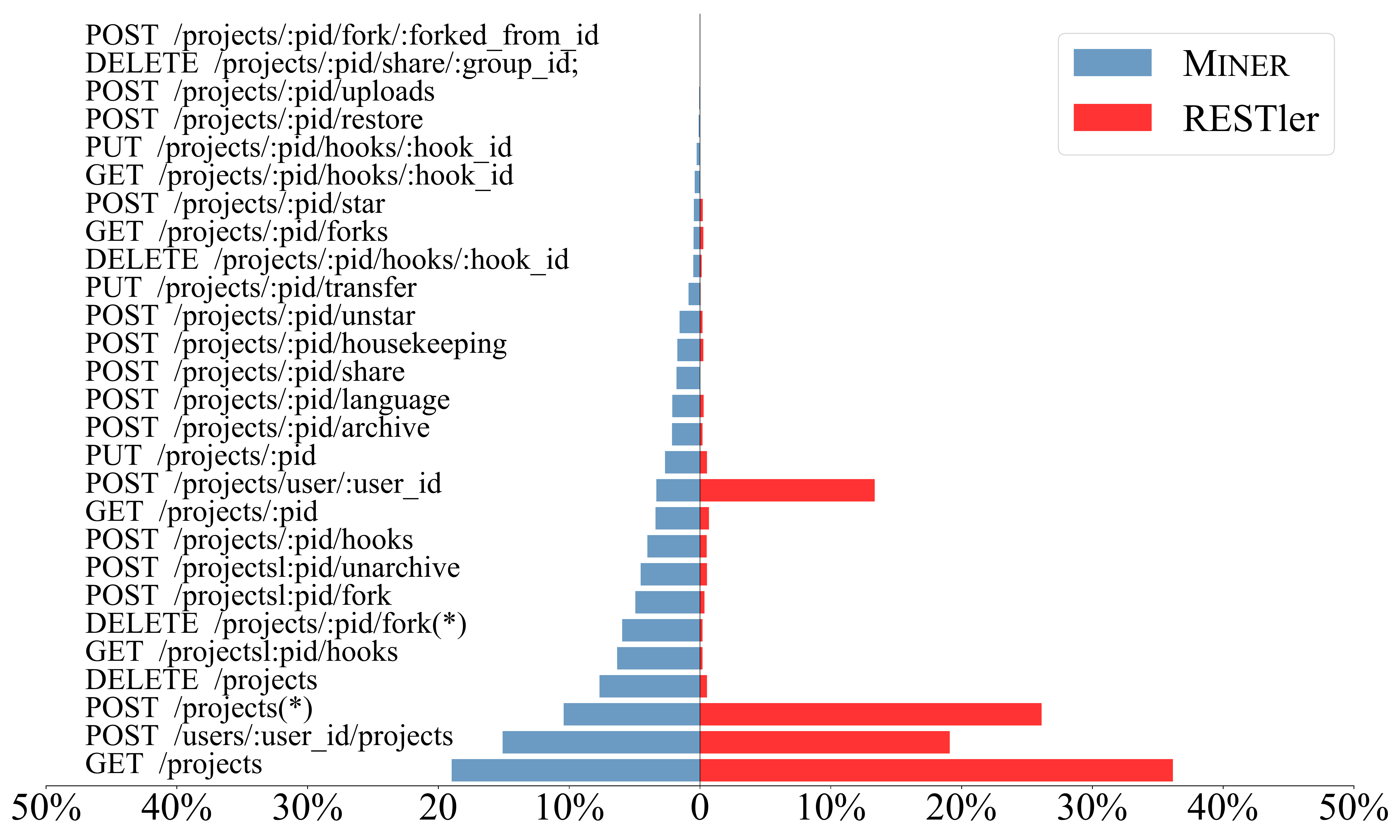}
            \setlength{\abovecaptionskip}{-0.4cm}
            \caption{GitLab Projects API.}
        \end{subfigure}
        \hfill
        \begin{subfigure}[b]{0.33\textwidth}
            \centering
            \includegraphics[width=2.3in,height=1.45in]{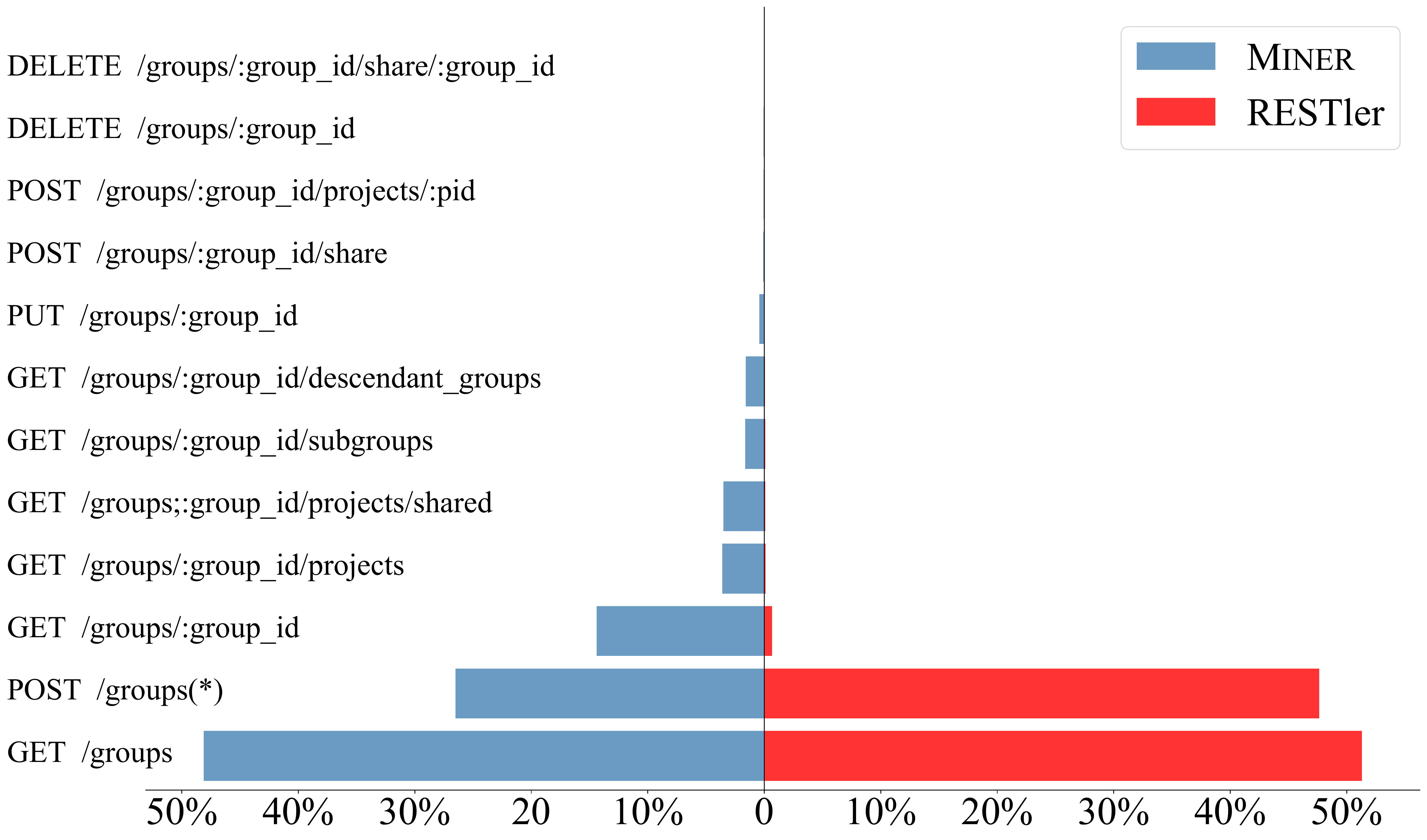}
            \setlength{\abovecaptionskip}{-0.4cm}
            \caption{GitLab Groups API.}
        \end{subfigure}
        \hfill
        \begin{subfigure}[b]{0.33\textwidth}
            \centering
            \includegraphics[width=2.3in,height=1.45in]{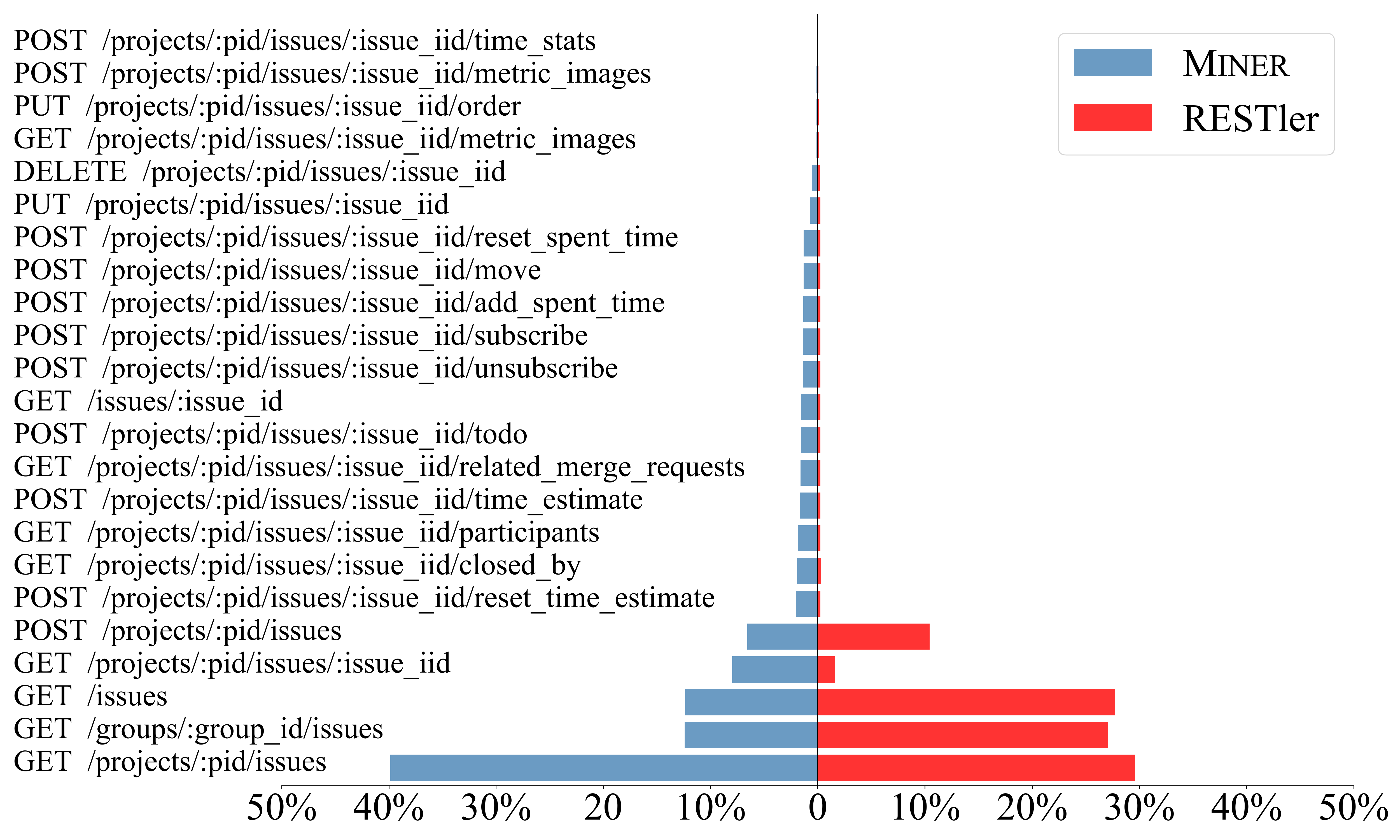}
            \setlength{\abovecaptionskip}{-0.4cm}
            \caption{GitLab Issues API.}
        \end{subfigure}
        \vspace{-0.9mm}
    \caption{The distribution of executions of different request templates sent by RESTler and \sysname, where the request template with $(*)$ means that it has multiple versions to provide different target object ids or it contains different parameters.} 
           \label{fig:executions}     
   \end{figure*}

$\bullet$
The pass rate of \sysnamepart slightly decreases when using a longer iteration duration to invoke the \texttt{Training Module}. 
For instance, the pass rate of \sysnamepart decreases from 90.90\% to 90.23\% when using larger iteration durations to fuzz GitLab Issues API. 
We infer the reasons for the results as follows. 
The slight difference in pass rate is caused by the request generation in the first period, 
in which \sysnamepart does not have the \textit{param-value lists} generated by the attention model and 
uses the traditional generation method to generate requests with a low pass rate. 
Thus, the larger the iteration duration is, the more times \sysnamepart generates requests with a low pass rate in the first period. 
After the first time invoking the \texttt{Training Module} 
and obtaining \textit{param-value lists}, 
\sysnamepart improves the generation quality of requests and significantly increases the pass rate. 
As a result, when using 3 hours as the iteration duration compared to 1 and 2 hours, 
\sysnamepart generates more requests with the traditional generation method, which results in a slightly lower average pass rate.

\subsection{Execution Distribution of Requests}\label{sec:distributionrequest}

In this subsection, we conduct the analysis to 
count the executions of different request templates sent by RESTler and \sysname. 
Each evaluation lasts for 12 hours, and 
the distribution of the executions of different request templates 
is shown in Fig.~\ref{fig:executions}, 
from which we have the following conclusion. 
{\sysname assigns more executions to diverse request templates than RESTler.}  
For instance, as shown in Fig.~\ref{fig:executions}, 
\sysname assigns most executions to 17 templates of requests when fuzzing GitLab Projects API. 
On the contrary, RESTler mainly tests GitLab with 4 request templates. 
When fuzzing GitLab Groups API, 
\sysname  assigns significantly more executions to 5 extra request templates compared to RESTler. 
Therefore, based on more different request templates, 
\sysname can  generate more diverse requests with different parameter values  compared to RESTler. 
Then, \sysname triggers more kinds of behaviors of a target cloud service with these requests, which can explore more and deeper states. 
Thus, the aforementioned conclusion can serve as one of the reasons 
that \sysname performs better than RESTler on line coverage and error discovery.

\section{Discussion and  Limitation}\label{sec:discussion}

\vspace{-1mm}
\subsection{Data-Driven Sequence Generation}
In this paper, \sysname utilizes the attention model to learn 
and generate \emph{param-value lists} for each request, 
which improves the request generation quality. 
In addition, 
the model can also be used to explore the implicit relations across different requests 
and provide key mutation strategies for a request sequence. 
Due to not having enough training data collected in the fuzzing process, 
it is hard for us to train an attention model for sequence generation. 
However, there can be sufficient training data for sequence generation 
in different 
scenarios like continuous fuzzing and parallel fuzzing. 
Therefore, utilizing a machine learning model to provide key mutation strategies for a request sequence 
can be an interesting topic in future work. 

\vspace{-0.2cm}
\subsection{Error Reproduction Across Sequences}\label{sec:hardreproduce}

Due to the change in the server states, some unique errors can only be triggered in a specific state of a cloud service, while cannot be reproduced in future analysis. It is a common problem in error reproduction for REST API fuzzing. For instance, a fuzzer can trigger an error by accessing a resource, which is created a few hours ago by other request sequences. However, the fuzzer deletes this resource in the following fuzzing process, making the error irreproducible. Thus, how to analyze the correlation between request sequences that are far apart and how to reproduce this kind of errors can be an interesting research direction for REST API fuzzing. 

\section{Related Work}\label{sec:related}

 \subsection{REST API Fuzzing}
 In 2019, Atlidakis et al. presented the first stateful REST API fuzzer named \emph{RESTler} to automatically fuzz a cloud service via its REST API~\cite{restler}. 
 Then, Atlidakis et al. implemented several security rule checkers in RESTler 
that can automatically detect violations of these security rules~\cite{atlidakis2020checking}.
 Godefroid et al. defined differential regression testing for REST APIs, 
 which leveraged RESTler to construct network logs on different versions of the target REST APIs 
 and detected service and specification regressions by comparing these network logs~\cite{godefroid2020differential}. 
 Godefroid et al. also studied how to generate data payloads learned from REST API specifications 
 and found data-processing bugs in cloud services~\cite{godefroid2020intelligent}. 
 Based on RESTler, Pythia is presented to leverage a machine learning model to 
 decide different mutation strategies on different positions of a request. 
 Pythia also leverages the code coverage information to guide the fuzzing process, 
 which needs manual cost before fuzzing~\cite{atlidakis2020pythia}.

 Different from existing REST API fuzzers, \sysname utilizes an attention model to locate the parameters to be mutated and provide the appropriate values, which is implemented by using the predicted \emph{param-value lists} in request generation. 
 Furthermore, \sysname leverages the other two data-driven designs to increase the execution times on long request sequences and explore incorrect parameter usage errors. 

 \subsection{Generation-based Fuzzing}
 Similar to REST API fuzzing,  
 most generation-based fuzzers generate  test cases 
 with specific input formats to fuzz a target program with syntax and semantic checking~\cite{godefroid2008grammar, holler2012fuzzing, dewey2014language, jsfunfuzz, wang2021mpinspector}. 
 Wang et al. presented a 
 novel data-driven seed generation approach named \emph{Skyfire} 
 to learn the syntax and semantic rules and generate the inputs with specific input formats,   
 which are used as the initial seeds for fuzzers~\cite{wang2017skyfire}. 
 Han et al. presented 
 a novel generation algorithm named \emph{semantics-aware assembly}  
to generate test cases 
 with semantical and syntactical correctness~\cite{han2019codealchemist}. 
\emph{Nautilus} combines the usage of grammars 
with code coverage feedback, 
which generates the test cases with a higher probability of having semantical and syntactical correctness~\cite{aschermann2019nautilus}. 
Lee et al. presented a neural network language model-guided fuzzer 
named \emph{Montage} to explore JavaScript engine vulnerabilities~\cite{lee2020montage}. 
 \emph{Fuzzilli} 
 utilizes a designed Intermediate Representation (IR) to build syntactically and semantically correct test cases~\cite{fuzzilli}.


  \subsection{Machine Learning-based Fuzzing}
 Multiple works  focus on improving fuzzers
with machine learning techniques~\cite{lv2018smartseed, godefroid2017learn, sivakorn2017hvlearn, 
nichols2017faster,  bottinger2018deep, rajpal2017not,lyu2022slime}. 
 \emph{NEUZZ} employs a neural network model to learn 
 the real-world program's branching behaviors, 
 and then utilizes the program smoothing technique to locate the bytes 
 in a test case that influence the branching behaviors~\cite{she2019neuzz}. 
 \textit{MTFuzz} presented by She et al. utilizes a multi-task neural network to  
 learn a compact embedding of program input spaces. 
  The neural network guides the mutation process by predicting which input bytes with the highest likelihood to impact code
 coverage~\cite{she2020mtfuzz}. 

\subsection{Mutation-based Fuzzing}
Mutation-based fuzzing focuses on exploring unique bugs on a target program without the input format requirement, 
which is different from REST API fuzzing. 
However, the following development directions of mutation-based fuzzing can still 
inspire future  REST API fuzzing.

Several studies make use of symbolic execution~\cite{stephens2016driller,yun2018qsym,zhao2019send,cho2019intriguer,kim2020hfl,huang2020pangolin}
to solve difficult path constraints and improve code coverage. 
Profuzzer includes a lightweight mechanism to discover the relationship between input bytes and program behaviors~\cite{you2019profuzzer}.  
 Aschermann et al. presented \emph{Redqueen} to solve magic bytes and checksums 
 automatically with input-to-state correspondence~\cite{aschermann2019redqueen}. 
Gan et al. presented a lightweight data flow-sensitive fuzzing
solution named \emph{GREYONE}, which contains  a taint-guided mutation strategy to mutate a seed on appropriate positions with appropriate values  
and a conformance-guided evolution solution to collect better seeds in the queue~\cite{gan2020greyone}.

Energy allocation strategies used in mutation-based fuzzing assign more energy to explore low-frequency paths, untouched branches and effective mutations~\cite{AFLFast,yueecofuzz,collafl,lyu2019mopt,lyu2022ems}. 
Several studies allocate energy based on multi-objective metrics~\cite{zhao2021moofuzz,alshahwan2018deploying,mao2016sapienz}. 
\textit{AFL++HIER} 
leverages a multi-armed bandit model to allocate energy to different clusters of seeds with multi-level coverage
metrics~\cite{wang2021aflhier}.

Inspired by these state-of-the-art solutions, in future work, 
REST API fuzzing can be further developed 
by 1) locating request templates that are most relevant to error discovery 
and 2) assigning more mutation energy to the request sequences with better error discovery and code coverage.

\section{Conclusion}\label{sec:conclusion}

To solve the limitations of REST API fuzzers and improve their error discovery performance, 
we present a hybrid data-driven approach 
with three new designs, 
which help find security bugs triggered by long request sequences and explore incorrect parameter usage errors. 
We implement the prototype \sysname based on our approach, 
and evaluate \sysname against 
RESTler 
on 3 open-sourced cloud services via 11 REST APIs. 
The results show that \sysname performs much better than RESTler 
on request generation and error discovery. 
Based on manual analysis, \sysname also finds 10 more real errors than RESTler, 
including 5 security bugs that try to access deleted resources. 
Furthermore, using the published bugs of GitLab as the ground truth, 
we demonstrate 
the significant performance of \sysname on serious bug discovery. 
In addition, we conduct extensive analysis to 
demonstrate the outstanding performance of \sysname 
on the sequence extension, line coverage and time overhead. 
Overall, 
our approach can serve as a new direction 
to improve the sequence extension, pass rate, and error discovery of REST API fuzzers.

\section*{Acknowledgments}

We sincerely appreciate the guidance from the shepherd. 
We would also  like to thank the anonymous reviewers for their valuable comments and input to improve our paper. 
This work was partly supported by NSFC under No. U1936215, the State Key Laboratory of Computer Architecture (ICT, CAS) under Grant No. CARCHA202001, and the Fundamental Research Funds for the Central Universities (Zhejiang University NGICS Platform). 



\bibliographystyle{plain}
\bibliography{00main}

\begin{thebibliography}{10}

\bibitem{apifuzzer}
{APIFuzzer: HTTP API Testing Framework}.
\newblock \url{https://github.com/KissPeter/APIFuzzer}.

\bibitem{bugzilla}
{Bugzilla}.
\newblock \url{https://www.bugzilla.org}.

\bibitem{coverband}
{Coverband}.
\newblock \url{https://github.com/danmayer/coverband}.

\bibitem{jsfunfuzz}
Funfuzz.
\newblock \url{https://github.com/MozillaSecurity/funfuzz}.

\bibitem{gitlab}
{GitLab}.
\newblock \url{https://gitlab.com/gitlab-org/gitlab}.

\bibitem{qualys}
{Qualys Web Application Scanning (WAS)}.
\newblock \url{https://www.qualys.com/apps/web-app-scanning}.

\bibitem{tntfuzzer}
{TnT-Fuzzer: OpenAPI Fuzzer Written in Python}.
\newblock \url{https://github.com/Teebytes/TnT-Fuzzer}.

\bibitem{wordpress}
{WordPress}.
\newblock \url{https://wordpress.org}.

\bibitem{alshahwan2018deploying}
Nadia Alshahwan, Xinbo Gao, Mark Harman, Yue Jia, Ke~Mao, Alexander Mols,
  Taijin Tei, and Ilya Zorin.
\newblock {Deploying Search based Software Engineering with Sapienz at
  Facebook}.
\newblock In {\em International Symposium on Search Based Software
  Engineering}, pages 3--45, 2018.

\bibitem{aschermann2019nautilus}
Cornelius Aschermann, Tommaso Frassetto, Thorsten Holz, Patrick Jauernig,
  Ahmad-Reza Sadeghi, and Daniel Teuchert.
\newblock {NAUTILUS: Fishing for Deep Bugs with Grammars}.
\newblock In {\em Network and Distributed Systems Security Symposium}, 2019.

\bibitem{aschermann2019redqueen}
Cornelius Aschermann, Sergej Schumilo, Tim Blazytko, Robert Gawlik, and
  Thorsten Holz.
\newblock {Redqueen: Fuzzing with Input-to-State Correspondence}.
\newblock In {\em Network and Distributed Systems Security Symposium}, 2019.

\bibitem{restler}
V.~Atlidakis, P.~Godefroid, and M.~Polishchuk.
\newblock {RESTler: Stateful REST API Fuzzing}.
\newblock In {\em Proceedings of the 41st International Conference on Software
  Engineering}, pages 748--758, 2019.

\bibitem{atlidakis2020pythia}
Vaggelis Atlidakis, Roxana Geambasu, Patrice Godefroid, Marina Polishchuk, and
  Baishakhi Ray.
\newblock {Pythia: Grammar-based Fuzzing of REST APIs with Coverage-guided
  Feedback and Learning-based Mutations}.
\newblock {\em arXiv preprint:2005.11498}, 2020.

\bibitem{atlidakis2020checking}
Vaggelis Atlidakis, Patrice Godefroid, and Marina Polishchunk.
\newblock {Checking Security Properties of Cloud Service REST APIs}.
\newblock In {\em 2020 IEEE 13th International Conference on Software Testing,
  Validation and Verification}, pages 387--397, 2020.

\bibitem{AFLFast}
Marcel B{\"o}hme, Van-Thuan Pham, and Abhik Roychoudhury.
\newblock {Coverage-based Greybox Fuzzing as Markov Chain}.
\newblock In {\em ACM SIGSAC Conference on Computer and Communications
  Security}, pages 1032--1043, 2016.

\bibitem{bottinger2018deep}
Konstantin B{\"o}ttinger, Patrice Godefroid, and Rishabh Singh.
\newblock {Deep Reinforcement Fuzzing}.
\newblock In {\em 2018 IEEE Security and Privacy Workshops}, pages 116--122,
  2018.

\bibitem{cho2014learning}
Kyunghyun Cho, Bart Van~Merri{\"e}nboer, Caglar Gulcehre, Dzmitry Bahdanau,
  Fethi Bougares, Holger Schwenk, and Yoshua Bengio.
\newblock {Learning Phrase Representations using RNN Encoder–Decoder for
  Statistical Machine Translation}.
\newblock {\em arXiv preprint:1406.1078}, 2014.

\bibitem{cho2019intriguer}
Mingi Cho, Seoyoung Kim, and Taekyoung Kwon.
\newblock {Intriguer: Field-Level Constraint Solving for Hybrid Fuzzing}.
\newblock In {\em ACM SIGSAC Conference on Computer and Communications
  Security}, pages 515--530, 2019.

\bibitem{chung2014empirical}
Junyoung Chung, Caglar Gulcehre, KyungHyun Cho, and Yoshua Bengio.
\newblock {Empirical Evaluation of Gated Recurrent Neural Networks on Sequence
  Modeling}.
\newblock {\em arXiv preprint:1412.3555}, 2014.

\bibitem{devlin2018bert}
Jacob Devlin, Ming-Wei Chang, Kenton Lee, and Kristina Toutanova.
\newblock {BERT: Pre-training of Deep Bidirectional Transformers for Language
  Understanding}.
\newblock {\em arXiv preprint:1810.04805}, 2018.

\bibitem{dewey2014language}
Kyle Dewey, Jared Roesch, and Ben Hardekopf.
\newblock {Language Fuzzing Using Constraint Logic Programming}.
\newblock In {\em International Conference on Automated Software Engineering},
  pages 725--730, 2014.

\bibitem{fielding2000architectural}
Roy~Thomas Fielding.
\newblock {\em {Architectural Styles and the Design of Network-based Software
  Architectures}}.
\newblock University of California, Irvine, 2000.

\bibitem{gan2020greyone}
Shuitao Gan, Chao Zhang, Peng Chen, Bodong Zhao, Xiaojun Qin, Dong Wu, and
  Zuoning Chen.
\newblock {GREYONE: Data Flow Sensitive Fuzzing}.
\newblock In {\em USENIX Security Symposium}, pages 2577--2594, 2020.

\bibitem{collafl}
Shuitao Gan, Chao Zhang, Xiaojun Qin, Xuwen Tu, Kang Li, Zhongyu Pei, and
  Zuoning Chen.
\newblock {CollAFL: Path sensitive fuzzing}.
\newblock In {\em Symposium on Security and Privacy}, pages 679--696, 2018.

\bibitem{godefroid2020intelligent}
Patrice Godefroid, Bo-Yuan Huang, and Marina Polishchuk.
\newblock {Intelligent REST API Data Fuzzing}.
\newblock In {\em ACM Joint Meeting on European Software Engineering Conference
  and Symposium on the Foundations of Software Engineering}, pages 725--736,
  2020.

\bibitem{godefroid2008grammar}
Patrice Godefroid, Adam Kiezun, and Michael~Y Levin.
\newblock {Grammar-based Whitebox Fuzzing}.
\newblock In {\em ACM Sigplan Notices}, 2008.

\bibitem{godefroid2020differential}
Patrice Godefroid, Daniel Lehmann, and Marina Polishchuk.
\newblock {Differential Regression Testing for REST APIs}.
\newblock In {\em International Symposium on Software Testing and Analysis},
  pages 312--323, 2020.

\bibitem{godefroid2017learn}
Patrice Godefroid, Hila Peleg, and Rishabh Singh.
\newblock {Learn\&Fuzz: Machine Learning for Input Fuzzing}.
\newblock In {\em International Conference on Automated Software Engineering},
  pages 50--59, 2017.

\bibitem{fuzzilli}
Samuel Gro{\ss}.
\newblock {FuzzIL: Coverage Guided Fuzzing for JavaScript Engines}.
\newblock {\em Department of Informatics, Karlsruhe Institute of Technology},
  2018.

\bibitem{han2019codealchemist}
HyungSeok Han, DongHyeon Oh, and Sang~Kil Cha.
\newblock {CodeAlchemist: Semantics-Aware Code Generation to Find
  Vulnerabilities in JavaScript Engines}.
\newblock In {\em Network and Distributed Systems Security Symposium}, 2019.

\bibitem{hochreiter1997long}
Sepp Hochreiter and J{\"u}rgen Schmidhuber.
\newblock {Long Short-Term Memory}.
\newblock {\em Neural computation}, 9:1735--1780, 1997.

\bibitem{holler2012fuzzing}
Christian Holler, Kim Herzig, and Andreas Zeller.
\newblock {Fuzzing with Code Fragments}.
\newblock In {\em USENIX Security Symposium}, pages 445--458, 2012.

\bibitem{huang2020pangolin}
Heqing Huang, Peisen Yao, Rongxin Wu, Qingkai Shi, and Charles Zhang.
\newblock {Pangolin: Incremental Hybrid Fuzzing with Polyhedral Path
  Abstraction}.
\newblock In {\em Symposium on Security and Privacy}, pages 1613--1627, 2020.

\bibitem{kim2020hfl}
Kyungtae Kim, Dae~R Jeong, Chung~Hwan Kim, Yeongjin Jang, Insik Shin, and
  Byoungyoung Lee.
\newblock {HFL: Hybrid Fuzzing on the Linux Kernel}.
\newblock In {\em Network and Distributed Systems Security Symposium}, 2020.

\bibitem{lee2020montage}
Suyoung Lee, HyungSeok Han, Sang~Kil Cha, and Sooel Son.
\newblock {Montage: A Neural Network Language Model-Guided JavaScript Engine
  Fuzzer}.
\newblock In {\em USENIX Security Symposium}, pages 2613--2630, 2020.

\bibitem{luong2015effective}
Minh-Thang Luong, Hieu Pham, and Christopher~D Manning.
\newblock {Effective Approaches to Attention-based Neural Machine Translation}.
\newblock {\em arXiv preprint:1508.04025}, 2015.

\bibitem{lv2018smartseed}
Chenyang Lyu, Shouling Ji, Yuwei Li, Junfeng Zhou, Jianhai Chen, Pan Zhou, and
  Jing Chen.
\newblock {SmartSeed: Smart Seed Generation for Efficient Fuzzing}.
\newblock {\em arXiv preprint:1807.02606}, 2018.

\bibitem{lyu2019mopt}
Chenyang Lyu, Shouling Ji, Chao Zhang, Yuwei Li, Wei-Han Lee, Yu~Song, and
  Raheem Beyah.
\newblock {MOPT: Optimized Mutation Scheduling for Fuzzers}.
\newblock In {\em USENIX Security Symposium}, pages 1949--1966, 2019.

\bibitem{lyu2022ems}
Chenyang Lyu, Shouling Ji, Xuhong Zhang, Hong Liang, Binbin Zhao, Kangjie Lu,
  and Raheem Beyah.
\newblock {EMS: History-Driven Mutation for Coverage-based Fuzzing}.
\newblock In {\em Network and Distributed Systems Security Symposium}, 2022.

\bibitem{lyu2022slime}
Chenyang Lyu, Hong Liang, Shouling Ji, Xuhong Zhang, Binbin Zhao, Meng Han, Yun
  Li, Zhe Wang, Wenhai Wang, and Raheem Beyah.
\newblock {SLIME: Program-Sensitive Energy Allocation for Fuzzing}.
\newblock In {\em International Symposium on Software Testing and Analysis},
  pages 365--377, 2022.

\bibitem{mao2016sapienz}
Ke~Mao, Mark Harman, and Yue Jia.
\newblock {Sapienz: Multi-objective Automated Testing for Android
  Applications}.
\newblock In {\em International Symposium on Software Testing and Analysis},
  pages 94--105, 2016.

\bibitem{nichols2017faster}
Nicole Nichols, Mark Raugas, Robert Jasper, and Nathan Hilliard.
\newblock {Faster Fuzzing: Reinitialization with Deep Neural Models}.
\newblock {\em arXiv preprint:1711.02807}, 2017.

\bibitem{rajpal2017not}
Mohit Rajpal, William Blum, and Rishabh Singh.
\newblock {Not All Bytes are Equal: Neural Byte Sieve for Fuzzing}.
\newblock {\em arXiv preprint:1711.04596}, 2017.

\bibitem{she2020mtfuzz}
Dongdong She, Rahul Krishna, Lu~Yan, Suman Jana, and Baishakhi Ray.
\newblock {MTFuzz: Fuzzing with a Multi-task Neural Network}.
\newblock In {\em ACM Joint Meeting on European Software Engineering Conference
  and Symposium on the Foundations of Software Engineering}, pages 737--749,
  2020.

\bibitem{she2019neuzz}
Dongdong She, Kexin Pei, Dave Epstein, Junfeng Yang, Baishakhi Ray, and Suman
  Jana.
\newblock {NEUZZ: Efficient Fuzzing with Neural Program Smoothing}.
\newblock In {\em Symposium on Security and Privacy}, pages 803--817, 2019.

\bibitem{sivakorn2017hvlearn}
Suphannee Sivakorn, George Argyros, Kexin Pei, Angelos~D Keromytis, and Suman
  Jana.
\newblock {HVLearn: Automated Black-box Analysis of Hostname Verification in
  SSL/TLS Implementations}.
\newblock In {\em Symposium on Security and Privacy}, pages 521--538, 2017.

\bibitem{stephens2016driller}
Nick Stephens, John Grosen, Christopher Salls, Andrew Dutcher, Ruoyu Wang,
  Jacopo Corbetta, Yan Shoshitaishvili, Christopher Kruegel, and Giovanni
  Vigna.
\newblock {Driller: Augmenting Fuzzing Through Selective Symbolic Execution}.
\newblock In {\em Network and Distributed Systems Security Symposium}, 2016.

\bibitem{wang2021aflhier}
Jinghan Wang, Chengyu Song, and Heng Yin.
\newblock {Reinforcement Learning-based Hierarchical Seed Scheduling for
  Greybox Fuzzing}.
\newblock In {\em Network and Distributed Systems Security Symposium}, 2021.

\bibitem{wang2017skyfire}
Junjie Wang, Bihuan Chen, Lei Wei, and Yang Liu.
\newblock {Skyfire: Data-Driven Seed Generation for Fuzzing}.
\newblock In {\em Symposium on Security and Privacy}, pages 579--594, 2017.

\bibitem{wang2021mpinspector}
Qinying Wang, Shouling Ji, Yuan Tian, Xuhong Zhang, Binbin Zhao, Yuhong Kan,
  Zhaowei Lin, Changting Lin, Shuiguang Deng, and Alex~X Liu.
\newblock Mpinspector: a systematic and automatic approach for evaluating the
  security of iot messaging protocols.
\newblock In {\em USENIX Security Symposium}, pages 4205--4222, 2021.

\bibitem{you2019profuzzer}
Wei You, Xueqiang Wang, Shiqing Ma, Jianjun Huang, Xiangyu Zhang, XiaoFeng
  Wang, and Bin Liang.
\newblock {ProFuzzer: On-the-fly Input Type Probing for Better Zero-day
  Vulnerability Discovery}.
\newblock In {\em Symposium on Security and Privacy}, pages 769--786, 2019.

\bibitem{yueecofuzz}
Tai Yue, Pengfei Wang, Yong Tang, Enze Wang, Bo~Yu, Kai Lu, and Xu~Zhou.
\newblock {EcoFuzz: Adaptive Energy-Saving Greybox Fuzzing as a Variant of the
  Adversarial Multi-Armed Bandit}.
\newblock In {\em USENIX Security Symposium}, pages 2307--2324, 2020.

\bibitem{yun2018qsym}
Insu Yun, Sangho Lee, Meng Xu, Yeongjin Jang, and Taesoo Kim.
\newblock {QSYM: A Practical Concolic Execution Engine Tailored for Hybrid
  Fuzzing}.
\newblock In {\em USENIX Security Symposium}, pages 745--761, 2018.

\bibitem{zhao2019send}
Lei Zhao, Yue Duan, Heng Yin, and Jifeng Xuan.
\newblock {Send Hardest Problems My Way: Probabilistic Path Prioritization for
  Hybrid Fuzzing}.
\newblock In {\em Network and Distributed Systems Security Symposium}, 2019.

\bibitem{zhao2021moofuzz}
Xiaoqi Zhao, Haipeng Qu, Wenjie Lv, Shuo Li, and Jianliang Xu.
\newblock {MooFuzz: Many-Objective Optimization Seed Schedule for Fuzzer}.
\newblock {\em Mathematics}, 9:205, 2021.

\end{thebibliography}

\end{document}